\documentclass[aps,prd,twoside,twocolumn,nofootinbib,showpacs,floatfix]{revtex4-1}
\usepackage{amssymb}
\usepackage{amsmath,amssymb}
\usepackage{graphicx,bm}
\usepackage{slashed}
\usepackage{times}
\usepackage{epstopdf}
\usepackage{ulem} 
\usepackage[usenames]{color}
\usepackage{float}
\usepackage{subfigure}
\usepackage{rotating}
\usepackage{color}
\usepackage{multirow}
\usepackage{dcolumn}
\usepackage{overpic}
\usepackage{booktabs}
\usepackage{appendix}
\usepackage{makecell}
\usepackage{diagbox}

\renewcommand\sout{\bgroup \color{red} \ULdepth=-.5ex \ULset}

\usepackage[colorlinks, citecolor=blue,anchorcolor=red,menucolor=red,linkcolor=blue,filecolor=red,runcolor=red,urlcolor=blue,frenchlinks=red]{hyperref}
\begin{document}
\title{Flavor-spin symmetry of the $P^N_{\psi}/H_{\Omega_{ccc}}^N$ and $P^\Lambda_{\psi s}/H^{\Lambda}_{\Omega_{ccc}s}$ molecular states}
\author{Kan Chen}\email{chenk10@nwu.edu.cn}
\affiliation{School of Physics, Northwest University, Xi'an 710127, China}
\affiliation{Shaanxi Key Laboratory for theoretical Physics Frontiers, Xi'an 710127, China}
\affiliation{Institute of Modern Physics, Northwest University, Xi'an 710127, China}
\affiliation{Peng Huanwu Center for Fundamental Theory, Xi'an 710127, China}
\author{Bo Wang}\email{wangbo@hbu.edu.cn}
\affiliation{College of Physics Science \& Technology, Hebei University, Baoding 071002, China}
\affiliation{Hebei Key Laboratory of High-precision Computation and Application of Quantum Field Theory, Baoding, 071002, China}
\affiliation{Hebei Research Center of the Basic Discipline for Computational Physics, Baoding, 071002, China}

\begin{abstract}
Based on a contact lagrangian that incorporates the SU(3) flavor and SU(2) spin symmetries, we discuss the symmetry properties of the interactions among the heavy flavor meson-baryon $P_{\psi}^N$, $P_{\psi s}^\Lambda$ (with quark components [$n\bar{c}$][$nnc$], [$s\bar{c}$][$nnc$], or [$n\bar{c}$][$nsc$]) systems and di-baryon $H_{\Omega_{ccc}}^N$, $H^{\Lambda}_{\Omega_{ccc}s}$ (with quark components [$nnc$][$ncc$], [$nnc$][$scc$] or [$nsc$][$ncc$]) systems ($n=u$, $d$). The light quark components of the $P_{\psi}^N$ ($P_{\psi s}^\Lambda$) and $H_{\Omega_{ccc}}^N$ ($H_{\Omega_{ccc}s}^\Lambda$) systems have identical flavors, the interactions generated from the exchanges of light mesons in the $P_{\psi}^N$ ($P^\Lambda_{\psi s}$) systems should be very similar to that of the $H_{\Omega_{ccc}}^N$ ($H^{\Lambda}_{\Omega_{ccc}s}$) systems. We perform the single-channel and multi-channel calculations on the $P_{\psi}^N/P^\Lambda_{\psi s}/H_{\Omega_{ccc}}^N/H^{\Lambda}_{\Omega_{ccc}s}$ systems and introduce the SU(3) breaking effect to identify the different mass spectra among the $P_{\psi}^N$ ($H_{\Omega_{ccc}}^N$) and $P^\Lambda_{\psi s}$ ($H^{\Lambda}_{\Omega_{ccc}s}$) systems. We suggest two kinds of evidences for the existence of the flavor-spin symmetry among the heavy flavor $P_{\psi}^N/H_{\Omega_{ccc}}^N/P^\Lambda_{\psi s}/H^{\Lambda}_{\Omega_{ccc}s}$ molecule community, i.e., the mass arrangements of the $P_{\psi}^N/H_{\Omega_{ccc}}^N/P^\Lambda_{\psi s}/H^{\Lambda}_{\Omega_{ccc}s}$ mass spectra and the binding energies of the heavy flavor meson-baryon (di-baryon) systems attributed to the same contact potentials.
\end{abstract}
\maketitle

\vspace{2cm}
\section{Introduction}\label{Introduction}
In the past 10 years, many pentaquark candidates have been reported experimentally. The $P_{\psi}^N(4312)$, $P_{\psi}^N(4380)$, $P_{\psi}^N(4440)$, $P_{\psi}^N(4457)$ \cite{LHCb:2015yax,LHCb:2019kea}, $P_{\psi}^N(4337)$ \cite{LHCb:2021chn}, $P^\Lambda_{\psi s}(4338)$ \cite{LHCb:2022ogu}, and $P^\Lambda_{\psi s}(4459)$ \cite{LHCb:2020jpq}. Their masses are close to the thresholds of the $\Sigma_c^{(*)}\bar{D}^{(*)}$ or $\Xi_c\bar{D}^{(*)}$ systems, this general feature serves as an important evidence of their molecular nature. Besides the molecular interpretations, the compact pentaquark states, the hadro-charmounium states, the triangle singularities, and the cusp effects are also suggested to understand their underlying structures (Readers may refer to the reviews \cite{Chen:2016qju,Lebed:2016hpi,Esposito:2016noz,Hosaka:2016pey,Guo:2017jvc,Ali:2017jda,Liu:2019zoy,Brambilla:2019esw,Lucha:2021mwx,Chen:2021ftn,Chen:2022asf,Meng:2022ozq}) for more experimental and theoretical details).

In the molecular picture, as the components of the molecular state, the two hadrons bind together through the interactions that are mainly generated from the exchanges of light mesons. Thus, for the heavy flavor molecular states, the light quark component and the heavy quark component in each hadron play different roles. The non-relativistic heavy quark component in each hadron is stimulative to stabilize the heavy flavor molecular system, while the light quark component in one hadron and its correlation to the light quark component of another hadron will determine the types of the exchanged light mesons, and thus primarily determine the property of interaction among the two hadrons.

The $\Xi_{cc}^{(*)}$ baryon can be related to the anti-charmed $\bar{D}^{(*)}$ meson with the heavy diquark-antiquark symmetry (HDAS) \cite{Savage:1990di}. in other words, the existence of the molecular state that is below the $\Sigma_c^{(*)}\bar{D}^{(*)}$ or $\Xi_c\bar{D}^{(*)}$ threshold implies the existence of the molecular state that is below the $\Sigma_c^{(*)}\Xi_{cc}^{(*)}$ or $\Xi_c\Xi_{cc}^{(*)}$ threshold, respectively. In fact, due to the special role of heavy quark components in stabilizing the molecular system, the molecular state consists of the $\Sigma_c^{(*)}$ ($\Xi_c$) and $\Xi_{cc}^{(*)}$ components is expected to bind deeper than the molecular state consists of the $\Sigma_c^{(*)}$ ($\Xi_{c}$) and $\bar{D}^{(*)}$ components.

Similarly, the $\Omega_{cc}^{(*)}$ can be related to the anti-charmed-strange meson $\bar{D}_s^{(*)}$ with the HDAS. Note that the LHCb collaboration reported the $P^\Lambda_{\psi s}(4338)$ \cite{LHCb:2022ogu} state, which could be the good candidate of $\Xi_c\bar{D}$ molecular state. Besides, from the $J/\Psi\Lambda$ invariant spectrum, there might be a $P^\Lambda_{\psi s}(4255)$ structure near the $\Lambda_c\bar{D}_s$ threshold. The existence of $P^\Lambda_{\psi s}(4255)$ were discussed in some literatures \cite{Yan:2022wuz,Nakamura:2022gtu,Giachino:2022pws,Zhu:2022wpi,Chen:2022wkh,Feijoo:2022rxf}, since the threshold of the $\Lambda_c\bar{D}_s$ channel is very close to the $P^\Lambda_{\psi s}(4255)$, the $\Lambda_c\bar{D}_s$ interaction attracted much attentions. If the $P^\Lambda_{\psi s}(4255)$ do exist, we should also check the existence of its HDAS partner $\Lambda_c\Omega_{cc}$.

\begin{table}[htbp]
\renewcommand\arraystretch{1.5}
\caption{The heavy flavor meson-baryon and di-baryon systems and their corresponding thresholds. We adopt the isospin averaged masses for the single-charm mesons and baryons. For the $\Xi_{cc}$ baryon, we use the experimental mass from Refs. \cite{LHCb:2017iph,LHCb:2022rpd}. For the the doubly charmed baryons $\Xi_{cc}^*$ and $\Omega_{cc}^{(*)}$, we use their theoretical masses calculated from Ref. \cite{Ebert:2002ig}. All values are in units of MeV.}
\begin{tabular}{c|cccccccccccccc}
\toprule[0.8pt]
Quark content&\multicolumn{6}{c}{System and threshold}\\
\hline
\multirow{2}{*}{$(nnc)(n\bar{c})$}&$\Lambda_c\bar{D}$&$\Lambda_c\bar{D}^*$&$\Sigma_c\bar{D}$&$\Sigma_c^*\bar{D}$&$\Sigma_c\bar{D}^*$&$\Sigma_c^*\bar{D}^*$\\
                                  &$4153.7$          &$4295.0$             &$4320.8$        &$4385.4$           &$4462.1$           &4526.7 \\
\multirow{2}{*}{$(nnc)(ncc)$}&$\Lambda_c\Xi_{cc}$&$\Lambda_c\Xi_{cc}^*$&$\Sigma_c\Xi_{cc}$&$\Sigma^*_c\Xi_{cc}$&$\Sigma_c\Xi^*_{cc}$&$\Sigma^*_c\Xi^*_{cc}$\\
                             &$5907.9$           &$6013.5$             &6074.9            &$6139.5$            &$6180.5$            &$6245.1$\\
\hline
\multirow{2}{*}{$(nnc)(s\bar{c})$}&$\Lambda_c\bar{D}_s$&$\Lambda_c\bar{D}_s^*$&$\Sigma_c\bar{D}_s$&$\Sigma^*_c\bar{D}_s$&$\Sigma_c\bar{D}_s^*$&$\Sigma^*_c\bar{D}_s^*$\\
                                  &$4255.5$            &$4398.7$              &$4422.5$           &$4487.1$             &$4565.7$             &$4630.3$ \\
\multirow{2}{*}{$(nsc)(n\bar{c})$}&$\Xi_c\bar{D}$&$\Xi_c\bar{D}^*$&$\Xi_c^\prime \bar{D}$&$\Xi_c^* D$&$\Xi_c^\prime \bar{D}^*$&$\Xi_c^*\bar{D}^*$\\
                                  &$4336.7$      &$4478.0$        &$4446.0$              &$4513.2$   &$4587.4$                &$4654.5$\\
\multirow{2}{*}{$(nnc)(scc)$}&$\Lambda_c\Omega_{cc}$&$\Lambda_c\Omega_{cc}^*$&$\Sigma_c\Omega_{cc}$&$\Sigma_c^*\Omega_{cc}$&$\Sigma_c\Omega_{cc}^*$&$\Sigma_c^*\Omega_{cc}^*$\\
                             &$6064.5$              &$6158.5$                &$6231.5$             &$6296.1$               &$6325.5$               &$6390.1$\\
\multirow{2}{*}{$(nsc)(ncc)$}&$\Xi_c\Xi_{cc}$&$\Xi_{c}\Xi_{cc}^*$&$\Xi_c^\prime\Xi_{cc}$&$\Xi_{c}^*\Xi_{cc}$&$\Xi_{c}^\prime\Xi_{cc}^*$&$\Xi_c^*\Xi_{cc}^*$\\
                             &$6090.5$       &$6196.1$           &$6199.9$              &$6267.4$           &$6305.5$                  &$6373.0$\\
\hline
\bottomrule[0.8pt]
\end{tabular}\label{systems}
\end{table}

By replacing the $\bar{c}$ quark in $\bar{D}^{(*)}/\bar{D}_s^{(*)}$ mesons with the $cc$ pair, we obtain the expected hexaquark systems $H_{\Omega_{ccc}}^N/H^{\Lambda}_{\Omega_{ccc}s}$ that have identical light quark components to that of the $P_{\psi}^N/P^\Lambda_{\psi s}$ systems, we list the considered heavy flavor baryon-meson and di-baryon systems and their corresponding thresholds in Table \ref{systems}.

In Table \ref{systems}, we adopt the experimental mass of the $\Xi_{cc}^{++}$ baryon measured from the LHCb collaboration \cite{LHCb:2017iph,LHCb:2022rpd}. The mass spectrum of the doubly charmed baryons has been studied in many different frameworks, such as various quark models \cite{Roberts:2007ni,Migura:2006ep,Ebert:1996ec,Gerasyuta:1999pc,Itoh:2000um,Wang:2021rjk,Ebert:2002ig}, the bag models \cite{Fleck:1989mb,Ponce:1978gk,He:2004px}, Bethe-Salpeter equation \cite{Weng:2010rb}, Born-Oppenheimer EFT \cite{Soto:2020xpm}, Regge analysis \cite{Wei:2015gsa,Oudichhya:2021yln}, QCD sum rule \cite{Zhang:2008rt,Wang:2010it}, and lattice QCD \cite{Flynn:2003vz,Perez-Rubio:2015zqb,Bahtiyar:2020uuj}. We adopt the masses of the $\Xi_{cc}^{*}$ and $\Omega_{cc}^{(*)}$ doubly heavy baryons calculated from a relativistic quark model \cite{Ebert:2002ig}.

In this work, we collect the $P_{\psi}^N/P^\Lambda_{\psi s}/H_{\Omega_{ccc}}^N/H^{\Lambda}_{\Omega_{ccc}s}$ systems together and investigate the possible flavor-spin symmetry among the $P_{\psi}^N/P^\Lambda_{\psi s}/H_{\Omega_{ccc}}^N/H^{\Lambda}_{\Omega_{ccc}s}$ molecular community, i.e., the interactions of the molecular states can be related together through their flavor and spin structures. Our framework \cite{Chen:2021cfl,Chen:2021spf,Wang:2023hpp,Wang:2023eng} is based on the one-boson-exchange (OBE) picture at hadron level, the interaction of two hadrons is generated from the exchanges of light mesons. But we reexpress such process with a quark-level lagrangian possessing the SU(3) flavor symmetry and SU(2) spin symmetry. This framework is particularly convenient for the symmetry analysis of the interactions among different di-hadron systems \cite{Chen:2022wkh}.

The investigations on the triply charmed pentaquark or hexaquark states can be found in a few literatures. In Ref. \cite{Guo:2013xga}, the authors suggested that if the $X(3872)$ and $Z_b(10650)$ turn out to be the $D\bar{D}^*$ and $B^*\bar{B}^*$ bound states, then the HDAS implies the existences of the triply charmed and triply bottomed pentaquark states, respectively. Within the framework of OBE model, the triply heavy pentaquark systems and the triply charmed hexaquark states have also been investigated in Refs. \cite{Chen:2017jjn,Asanuma:2023atv} and in Ref. \cite{Chen:2018pzd}, respectively. By checking the result of the heavy dibaryon system $\Sigma_c\Xi_{cc}$ from lattice QCD \cite{Junnarkar:2019equ}, the authors in Ref. \cite{Pan:2019skd} proposed a model independent way to determine the spins of the $P_{\psi}^N(4440)$ and $P_{\psi}^N(4457)$, i.e., the mass arrangement of the $P_{\psi}^N$ bound state spectrum is related to the mass arrangement of the $H_{\Omega_{ccc}}^N$ bound state spectrum through the heavy quark spin symmetry. In this work, we will give an extended discussion on the symmetry properties of the interactions in the $P_{\psi}^N$/$H_{\Omega_{ccc}}^N$ as well as the $P^\Lambda_{\psi s}$/$H^{\Lambda}_{\Omega_{ccc}s}$ systems.

This paper is organized as follows. In Sec. \ref{framework}, we present our theoretical framework. In Sec. \ref{parameters and SC}, we discuss the way we determine the parameters introduced in our model, then we present how the flavor-spin symmetry manifests itself from our single-channel calculations on the $P_{\psi}^N/H_{\Omega_{ccc}}^N/P^\Lambda_{\psi s}/H^{\Lambda}_{\Omega_{ccc}s}$ systems. In Sec. \ref{MC}, we present and discuss the results from our multi-channel calculations on the $P_{\psi}^N/H_{\Omega_{ccc}}^N/P^\Lambda_{\psi s}/H^{\Lambda}_{\Omega_{ccc}s}$ systems. Sec. \ref{summary} is devoted to a summary.
\section{Framework}\label{framework}
In this section, we present our framework to calculate the mass spectra of the $P_{\psi}^N$/$H_{\Omega_{ccc}}^N$ and $P^\Lambda_{\psi s}$/$H^{\Lambda}_{\Omega_{ccc}s}$ states.
We introduce \cite{Chen:2021cfl,Chen:2021spf} the $S$ wave contact interactions via exchanging scalar and axial-vector light mesons
to collectively describe the interactions of the $P_{\psi}^N$/$H_{\Omega_{ccc}}^N$ and $P^\Lambda_{\psi s}$/$H^{\Lambda}_{\Omega_{ccc}s}$ systems
\begin{eqnarray}
\mathcal{L}=g_s\bar{q}\mathcal{S}q+g_a\bar{q}\gamma_{\mu}\gamma^5\mathcal{A}^{\mu}q.
\end{eqnarray}
The fictitious scalar field $\mathcal{S}$ and axial-vector field $\mathcal{A}^\mu$
can be expanded as
\begin{eqnarray}
\mathcal{S}=\mathcal{S}_3\lambda^i+\mathcal{S}_2\lambda^j+\mathcal{S}_1\lambda^8,\\
\mathcal{A}^{\mu}=\mathcal{A}^\mu_3\lambda^i+\mathcal{A}^\mu_2\lambda^j+\mathcal{A}^\mu_1\lambda^8,
\end{eqnarray}
respectively. The $\lambda^i$ ($i=1$, 2, 3), $\lambda^j$ ($j=4$, 5, 6, 7), and $\lambda^8$ are the generators of SU(3) group. $\mathcal{S}_3$ ($\mathcal{A}^\mu_3$), $\mathcal{S}_2$ ($\mathcal{A}^\mu_2$), and $\mathcal{S}_1$ ($\mathcal{A}^\mu_1$) denote the isospin triplet, isospin doublet, and isospin single scalar (axial-vector) fields, respectively.

The effective potential introduced from the exchanges of scalar and axial-vector mesons can be written as
\begin{eqnarray}
V=\tilde{g}_s\bm{\lambda}_1\cdot\bm{\lambda}_2+\tilde{g}_a\bm{\lambda}_1\cdot\bm{\lambda}_2\bm{\sigma}_1\cdot\bm{\sigma}_2.\label{potential1}
\end{eqnarray}
The $\bm{\sigma}_{1(2)}$ is the Pauli matrix in the spin space. The redefined coupling constants are $\tilde{g}_s=g_s^2/m_{\mathcal{S}}^2$ and $\tilde{g}_a=g_a^2/m_{\mathcal{A}}^2$, where $m_{\mathcal{S}}$ and $m_{\mathcal{A}}$ are the masses of the scalar and axial-vector light mesons.

To calculate the effective potential of the considered $P_{\psi}^N/P^\Lambda_{\psi s}/H_{\Omega_{ccc}}^N/H^{\Lambda}_{\Omega_{ccc}s}$ systems, we need to construct the wave functions of the mesons and baryons involved in Table \ref{systems}. We collectively present their flavor and spin wave functions in Table \ref{flavor WF} and \ref{spin WF}, respectively.
\begin{table}[htbp]
\setlength\tabcolsep{0.9pt} \caption{The flavor wave functions of the heavy flavor hadrons considered in this work. \label{flavor WF}}
\renewcommand\arraystretch{1.5}
\begin{tabular}{llc|llc}
\toprule[1pt]
Meson&$\left|Im_I\right\rangle$&$\phi^{M}_{Im_I}$&Meson&$\left|Im_I\right\rangle$&$\phi^{M}_{Im_I}$\\
\hline
$\bar{D}^{(*)0}$&$\left|\frac{1}{2}\frac{1}{2}\right\rangle$&$u\bar{c}$&$\bar{D}^{(*)-}$&$\left|\frac{1}{2}-\frac{1}{2}\right\rangle$&$d\bar{c}$\\
$\bar{D}_s^{(*)-}$&$\left|00\right\rangle$&$s\bar{c}$&&&\\
\hline
Baryon&$\left|Im_I\right\rangle$&$\phi^{B}_{Im_I}$&Baryon&$\left|Im_I\right\rangle$&$\phi^{B}_{Im_I}$\\
\hline
$\Lambda_c^+$&$\left|00\right\rangle$&$\frac{1}{\sqrt{2}}\left(du-ud\right)c$&$\Sigma_c^{(*)++}$&$\left|11\right\rangle$&$uuc$\\
$\Sigma_c^{(*)+}$&$\left|10\right\rangle$&$\frac{1}{\sqrt{2}}\left(ud+du\right)c$&$\Sigma_c^{(*)0}$&$\left|1-1\right\rangle$&$ddc$\\
$\Xi_c^+$&$\left|\frac{1}{2}\frac{1}{2}\right\rangle$&$\frac{1}{\sqrt{2}}\left(us-su\right)c$&$\Xi_c^0$&$\left|\frac{1}{2}-\frac{1}{2}\right\rangle$&$\frac{1}{\sqrt{2}}\left(ds-sd\right)c$\\
$\Xi_c^{\prime(*)+}$&$\left|\frac{1}{2}\frac{1}{2}\right\rangle$&$\frac{1}{\sqrt{2}}\left(us+su\right)c$&$\Xi_c^{\prime(*)0}$&$\left|\frac{1}{2}-\frac{1}{2}\right\rangle$&$\frac{1}{\sqrt{2}}\left(ds+sd\right)c$\\
$\Xi_{cc}^{(*)++}$&$|\frac{1}{2}\frac{1}{2}\rangle$&$ucc$&$\Xi_{cc}^{(*)+}$&$|\frac{1}{2}-\frac{1}{2}\rangle$&$dcc$\\
$\Omega_{cc}^{(*)+}$&$|00\rangle$&$scc$\\
\bottomrule[1pt]
\end{tabular}
\end{table}

\begin{table*}[htbp]
\setlength\tabcolsep{0.9pt} \caption{The spin wave functions of the heavy flavor hadrons considered in this work. \label{spin WF}}
\renewcommand\arraystretch{1.5}
\begin{tabular}{|clc|clc|}
\toprule[1pt]
Hadron&$\left|Sm_S\right\rangle$&$\phi^{M}_{Sm_S}$&Hadron&$\left|Sm_S\right\rangle$&$\phi^{M}_{Sm_S}$\\
\hline
\multirow{3}{*}{$\bar{D}$/$\bar{D}_s$}&\multirow{3}{*}{$|00\rangle$}&\multirow{3}{*}{$\frac{1}{\sqrt{2}}(\uparrow\downarrow-\downarrow\uparrow)$}& \multirow{3}{*}{$\bar{D}^*$/$\bar{D}_s^*$}&$|11\rangle$&$\uparrow\uparrow$\\
&&&&$|10\rangle$&$\frac{1}{\sqrt{2}}(\uparrow\downarrow+\downarrow\uparrow)$\\
&&&&$|1-1\rangle$&$\downarrow\downarrow$\\
\hline
Hadron&$\left|Sm_S\right\rangle$&$\phi^{B}_{Sm_S}$&Hadron&$\left|Sm_S\right\rangle$&$\phi^{B}_{Sm_S}$\\
\hline
\multirow{2}{*}{$\Lambda_c$/$\Xi_c$}&$|\frac{1}{2}\frac{1}{2}\rangle$&$\frac{1}{\sqrt{2}}(\uparrow\downarrow-\downarrow\uparrow)\uparrow$&&&\\
&$|\frac{1}{2}-\frac{1}{2}\rangle$&$\frac{1}{\sqrt{2}}(\uparrow\downarrow-\downarrow\uparrow)\downarrow$&&$|\frac{3}{2}\frac{3}{2}\rangle$&$\uparrow\uparrow\uparrow$\\
\multirow{2}{*}{$\Sigma_c$/$\Xi^\prime_c$}&$|\frac{1}{2}\frac{1}{2}\rangle$&$-\frac{1}{\sqrt{6}}(\uparrow\downarrow+\downarrow\uparrow)\uparrow+\sqrt{\frac{2}{3}}\uparrow\uparrow\downarrow$
&$\Sigma_c^*$/$\Xi_c^*$&$|\frac{3}{2}\frac{1}{2}\rangle$&$\sqrt{\frac{1}{3}}(\uparrow\uparrow\downarrow+\uparrow\downarrow\uparrow+\downarrow\uparrow\uparrow)$\\
&$|\frac{1}{2}-\frac{1}{2}\rangle$&$\frac{1}{\sqrt{6}}(\uparrow\downarrow+\downarrow\uparrow)\downarrow-\sqrt{\frac{2}{3}}\downarrow\downarrow\uparrow$
&$/\Xi_{cc}^*$/$\Omega_{cc}^*$&$|\frac{3}{2}-\frac{1}{2}\rangle$&$\sqrt{\frac{1}{3}}(\uparrow\downarrow\downarrow+\downarrow\uparrow\downarrow+\downarrow\downarrow\uparrow)$\\
\multirow{2}{*}{$\Xi_{cc}$/$\Omega_{cc}$}&$|\frac{1}{2}\frac{1}{2}\rangle$&$-\frac{1}{\sqrt{6}}\uparrow(\uparrow\downarrow+\downarrow\uparrow)+\sqrt{\frac{2}{3}}\downarrow\uparrow\uparrow$
&&$|\frac{3}{2}-\frac{3}{2}\rangle$&$\downarrow\downarrow\downarrow$\\
&$|\frac{1}{2}-\frac{1}{2}\rangle$&$\frac{1}{\sqrt{6}}\downarrow(\uparrow\downarrow+\downarrow\uparrow)-\sqrt{\frac{2}{3}}\uparrow\downarrow\downarrow$&&&\\
\bottomrule[1pt]
\end{tabular}
\end{table*}
With the above preparations, the total wave function of the considered heavy flavor di-hadron system can be written as
\begin{eqnarray}
|[H_1H_2]_J^I\rangle&=&\sum_{m_{I_1}m_{I_2}}C_{I_1,m_{I_1};I_2,m_{I_2}}^{I,I_z}\phi^{H_1}_{I_1,m_{I_1}}\phi^{H_2}_{I_2,m_{I_2}}\nonumber\\
&&\otimes\sum_{m_{S_1},m_{S_2}}C_{S_1,m_{S_1},S_2,m_{S_2}}^{J,J_z}\phi_{S_1,m_{S_1}}^{H_1}\phi_{S_2,m_{S_2}}^{H_2}.\nonumber\\
\end{eqnarray}
Here, $H_{1(2)}$  stands for the considered heavy flavor meson or baryon. The $C_{I_1,m_{I_1};I_2,m_{I_2}}^{I,I_z}$ and $C_{S_1,m_{S_1},S_2,m_{S_2}}^{J,J_z}$ are the Clebsch-Gordan coefficients.

With the constructed total wave function, the effective potential for a specific heavy flavor di-hadron system with total isospin $I$ and total angular momentum $J$ can be written as
\begin{eqnarray}
V_{[H_1H_2]_J^I}=\left\langle\left[H_1H_2\right]^I_J\left|V\right|\left[H_1H_2\right]^I_J\right\rangle.\label{potential}
\end{eqnarray}
Here, since we only consider the interactions that are introduced from the exchanges of light mesons, the operators $\bm{\lambda}_1\cdot\bm{\lambda}_2$ and $\bm{\lambda}_1\cdot\bm{\lambda}_2\bm{\sigma}_1\cdot\bm{\sigma}_2$ only act on the light quark components of the total wave function $|[H_1H_2]_J^I\rangle$.

At present, the experimentally observed $P_{\psi}^N$ \cite{LHCb:2015yax,LHCb:2019kea} and $P^\Lambda_{\psi s}$ \cite{LHCb:2022ogu,LHCb:2020jpq} molecular candidates have the lowest isospin $\frac{1}{2}$ and $0$, respectively. In Ref. \cite{Chen:2021cfl}, with the same framework, we proposed an isospin criterion to explain why the experimentally observed $P_{\psi}^N$/$P^\Lambda_{\psi s}$ molecular candidates prefer the lowest isospin numbers. This criterion suggest that when the light quark components of these systems couple to lower isospin numbers, the interactions generated from such configurations lead to more attractive forces, which is crucial for the formation of the heavy flavor bound states. Thus, in this work, we only focus on the $P_{\psi}^N/H_{\Omega_{ccc}}^N$ and $P^\Lambda_{\psi s}/H^{\Lambda}_{\Omega_{ccc}s}$ systems with the lowest isospin numbers, i.e., $I=1/2$ and $I=0$ for the $P_{\psi}^N/H_{\Omega_{ccc}}^N$ and $P^\Lambda_{\psi s}/H^{\Lambda}_{\Omega_{ccc}s}$ systems, respectively. 

We collect the considered channels with the lowest isospin and all possible total angular momentum numbers for each $P_{\psi}^N/P^\Lambda_{\psi s}/H_{\Omega_{ccc}}^N/H^{\Lambda}_{\Omega_{ccc}s}$ system  in Table \ref{coupled-channels}.
\begin{table}[htbp]
\setlength\tabcolsep{0.9pt} \caption{The heavy flavor meson-baryon channels ($P_{\psi}^N$ and $P^\Lambda_{\psi s}$ systems) and di-baryon channels ($H_{\Omega_{ccc}}^N$ and $H^{\Lambda}_{\Omega_{ccc}s}$ systems) with the lowest isospin numbers and all possible total angular momentum numbers. \label{coupled-channels}}
\renewcommand\arraystretch{1.5}
\begin{tabular}{|clc|clc|}
\toprule[1pt]
&$I(J^P)$&Channel\\
\hline
\multirow{3}{*}{$P_{\psi}^N$}&$\frac{1}{2}(\frac{1}{2}^-)$&$\Lambda_c\bar{D}$,$\Lambda_c\bar{D}^*$,$\Sigma_c\bar{D}$,$\Sigma_c\bar{D}^*$,$\Sigma_c^*\bar{D}^*$\\
                      &$\frac{1}{2}(\frac{3}{2}^-)$&$\Lambda_c\bar{D}^*$,$\Sigma_c^*\bar{D}$,$\Sigma_c\bar{D}^*$,$\Sigma_c^*\bar{D}^*$\\
                      &$\frac{1}{2}(\frac{5}{2}^-)$&$\Sigma^*_c\bar{D}^*$\\
                      \hline
\multirow{3}{*}{$P^{\Lambda}_{\psi s}$}&$0(\frac{1}{2}^-)$&$\Lambda_c\bar{D}_s$,$\Lambda_c\bar{D}^*_s$,$\Xi_c\bar{D}$,$\Xi_c\bar{D}^*$,$\Xi_c^\prime\bar{D}$,$\Xi_c^\prime\bar{D}^*$,$\Xi_c^*\bar{D}^*$\\
                         &$0(\frac{3}{2}^-)$&$\Lambda_c\bar{D}_s^*$,$\Xi_c\bar{D}^*$,$\Xi_c^*\bar{D}$,$\Xi_c^\prime\bar{D}^*$,$\Xi_c^*\bar{D}^*$\\
                         &$0(\frac{5}{2}^-)$&$\Xi_c^*\bar{D}^*$\\
                         \hline
\multirow{4}{*}{$H_{\Omega_{ccc}}^N$}&$\frac{1}{2}(0^+)$& $\Lambda_c\Xi_{cc}$, $\Sigma_c\Xi_{cc}$,$\Sigma_c^*\Xi_{cc}^*$\\
                      &$\frac{1}{2}(1^+)$& $\Lambda_c\Xi_{cc}$, $\Lambda_c\Xi_{cc}^*$, $\Sigma_c\Xi_{cc}$, $\Sigma_c^*\Xi_{cc}$, $\Sigma_c\Xi_{cc}^*$, $\Sigma_c^*\Xi_{cc}^*$\\                  &$\frac{1}{2}(2^+)$& $\Lambda_c\Xi_{cc}^*$,$\Sigma_c^*\Xi_{cc}$,$\Sigma_c\Xi_{cc}^*$,$\Sigma_c^*\Xi_{cc}^*$\\
                      &$\frac{1}{2}(3^+)$& $\Sigma_c^*\Xi_{cc}^*$\\
                      \hline
\multirow{4}{*}{$H_{\Omega_{ccc} s}^{\Lambda}$}&$\frac{1}{2}(0^+)$& $\Lambda_c\Omega_{cc}$, $\Xi_c\Xi_{cc}$,$\Xi_c^\prime\Xi_{cc}$,$\Xi_c^*\Xi_{cc}^*$\\
                      &$\frac{1}{2}(1^+)$& $\Lambda_c\Omega_{cc}^*$,$\Xi_c\Xi_{cc},\Xi_c\Xi_{cc}^*$,$\Xi_c^\prime\Xi_{cc}$,$\Xi_c^*\Xi_{cc}$,$\Xi_c^\prime\Xi_{cc}^*$,$\Xi_c^*\Xi_{cc}^*$\\
                      &$\frac{1}{2}(2^+)$& $\Xi_c\Xi_{cc}^*$,$\Xi_c^*\Xi_{cc}$,$\Xi_c^\prime\Xi_{cc}^*$,$\Xi_c^*\Xi_{cc}^*$\\
                      &$\frac{1}{2}(3^+)$& $\Xi_c^*\Xi_{cc}^*$\\
\bottomrule[1pt]
\end{tabular}
\end{table}

The effective potential matrices $\mathbb{V}_J^{P_{\psi}^N}$, $\mathbb{V}_J^{P^\Lambda_{\psi s}}$, $\mathbb{V}_J^{H_{\Omega_{ccc}}^N}$, $\mathbb{V}_J^{H^{\Lambda}_{\Omega_{ccc}s}}$ can be calculated from Eq. (\ref{potential}) with $J=1/2$, $3/2$, and $5/2$ for the $P_{\psi}^N$ and $P^\Lambda_{\psi s}$ systems, and with $J=0$, $1$, $2$, and $3$ for the $H_{\Omega_{ccc}}^N$ and $H^{\Lambda}_{\Omega_{ccc}s}$ systems. Then we find the bound state solutions by solving the following coupled-channel Lippmann-Schwinger equation
\begin{eqnarray}
\mathbb{T}\left(E\right)&=&\mathbb{V}+\mathbb{V}\mathbb{G}\left(E\right)\mathbb{T}\left(E\right),\label{LSE}
\end{eqnarray}
with
\begin{eqnarray}
\mathbb{V}=\left(
  \begin{array}{ccccc}
    v_{11} & \cdots & v_{1i} & \cdots & v_{1n} \\
    \vdots &  & \vdots &  & \vdots \\
    v_{j1} & \cdots & v_{ji} & \cdots & v_{jn} \\
    \vdots &  & \vdots &  & \vdots \\
    v_{n1} & \cdots & v_{ni} & \cdots & v_{nn} \\
  \end{array}
\right),
\end{eqnarray}
\begin{eqnarray}
\mathbb{T}(E)=\left(
  \begin{array}{ccccc}
    t_{11}(E) & \cdots & t_{1i}(E) & \cdots & t_{1n}(E) \\
    \vdots &  & \vdots &  & \vdots \\
    t_{j1}(E) & \cdots &t_{ji}(E) & \cdots & t_{jn}(E) \\
    \vdots &  & \vdots &  & \vdots \\
    t_{n1}(E) & \cdots & t_{ni}(E) & \cdots & t_{nn}(E) \\
  \end{array}
\right),
\end{eqnarray}
and
\begin{eqnarray}
\mathbb{G}(E)=\text{diag}\left\{G_{1}(E),\cdots,G_{i}(E),\cdots,G_{n}(E)\right\}.
\end{eqnarray}
Here, we adopt a dipole form factor $u(\Lambda)=(1+q^2/\Lambda^2)^{-2}$ \cite{Nakamura:2022gtu,Leinweber:2003dg,Wang:2007iw,Chen:2022wkh} to suppress the contributions from higher momenta, i.e.,
\begin{eqnarray}
G_{i}=\frac{1}{2\pi^2}\int dq \frac{q^2}{E-\sqrt{m_{i1}^2+q^2}-\sqrt{m_{i2}^2+q^2}}u^2(\Lambda).\nonumber\\
\end{eqnarray}
Three parameters $\tilde{g}_s$, $\tilde{g}_a$, and $\Lambda$ are introduced in our model, we will discuss the determination of these three parameters in Sec. \ref{parameters and SC}.

The bound state solution satisfies the following equation
\begin{eqnarray}
||\mathbb{I}-\mathbb{V}\mathbb{G}||=0.
\end{eqnarray}
The $\mathbb{I}$ is the unit matrix. For the bound state below the lowest channel, we find its solution in the first Riemann sheet of the lowest channel. For the quasi-bound state between the thresholds of the $i$-th (lower) and $j$-th (higher) channels, they generally has non-zero imaginary parts due to its non-trivial couplings to the lower $i$ channels. We find the quasi-bound state solution between the first Riemann sheet of the higher $j$-th channel and the second Riemann sheet of the lower $i$-th channel.

\section{Parameters and the results from our single-channel fromalism}\label{parameters and SC}
In this section, we briefly introduce the way we determine the three parameters $\tilde{g}_s$, $\tilde{g}_a$, and $\Lambda$ that are introduced in our model. Then we use a single channel formalism to give a preliminary discussion on how the flavor-spin symmetry manifests itself among the $P_{\psi}^N/P_{\psi s}^{\Lambda}/H_{\Omega_{ccc}}^N/H^{\Lambda}_{\Omega_{ccc}s}$ systems. 
\subsection{Determination of the parameters}\label{parameters}
We use the masses of the $P_{\psi}^N(4440)$ and $P_{\psi}^N(4457)$ \cite{LHCb:2019kea} as inputs to determine the $\tilde{g}_s$ and $\tilde{g}_a$. The $J^P$ numbers of these two states have not been measured yet, since they both lie slightly below the $\Sigma_c\bar{D}^*$ threshold, the following two scenarios are possible for their assignments
\begin{eqnarray}
&&\text{Scenario}\,1: P_{\psi}^N(4440)|\Sigma_c\bar{D}^*;\frac{1}{2}^-\rangle,
P_{\psi}^N(4457)|\Sigma_c\bar{D}^*;\frac{3}{2}^-\rangle,\nonumber\\\\
&&\text{Scenario}\,2: P_{\psi}^N(4457)|\Sigma_c\bar{D}^*;\frac{1}{2}^-\rangle,
P_{\psi}^N(4440)|\Sigma_c\bar{D}^*;\frac{3}{2}^-\rangle.\nonumber\\
\end{eqnarray}
We fix the cutoff $\Lambda$ at 1.0 GeV. Since we only consider the exchanges of the scalar and axial-vector light mesons, this value is comparable to the mass of the ground scalar or axial-vector mesons, and they are integrated out in our effective theory. Besides, we also checked that our numerical results have weak $\Lambda$-dependences around 1.0 GeV. With the above two sets of assignments, the $\tilde{g}_s$ and $\tilde{g}_a$ can be obtained by solving the following equations \cite{Chen:2022wkh}
\begin{eqnarray}
\text{Re}\left|\left|\mathbb{I}-\mathbb{V}_{1/2}^{P_{\psi}^{N}}\mathbb{G}_{1/2}^{P_{\psi}^{N}}\right|\right|=0,\label{M1}\\
\text{Im}\left|\left|\mathbb{I}-\mathbb{V}_{1/2}^{P_{\psi}^{N}}\mathbb{G}_{1/2}^{P_{\psi}^{N}}\right|\right|=0,\label{M2}\\
\text{Re}\left|\left|\mathbb{I}-\mathbb{V}_{3/2}^{P_{\psi}^{N}}\mathbb{G}_{3/2}^{P_{\psi}^{N}}\right|\right|=0,\label{M3}\\
\text{Im}\left|\left|\mathbb{I}-\mathbb{V}_{3/2}^{P_{\psi}^{N}}\mathbb{G}_{3/2}^{P_{\psi}^{N}}\right|\right|=0,\label{M4}
\end{eqnarray}
then the parameters $\tilde{g}_s$ and $\tilde{g}_a$ in scenario 1 and scenario 2 are obtained as
\begin{eqnarray}
&&\text{Scenario}\,1:\tilde{g}_s=8.28\,\text{GeV}^{-2}, \,\tilde{g}_a=-1.46\,\text{GeV}^{-2},\label{scenario 1}\\
&&\text{Scenario}\,2:\tilde{g}_s=9.12\,\text{GeV}^{-2},\,\tilde{g}_a=1.25\,\text{GeV}^{-2}.\label{scenario 2}
\end{eqnarray}
The parameter $\tilde{g}_s$ that is related to the operator $\bm{\lambda}_1\cdot\bm{\lambda}_2$ is much larger than the parameter $\tilde{g}_a$ that is related to the operator $\bm{\lambda}_1\cdot\bm{\lambda}_2\bm{\sigma}_1\cdot\bm{\sigma}_2$. The first term in Eq. (\ref{potential1}) provides the driving force for the formation of molecular states, the values of $\tilde{g}_a$ obtained from these two scenarios have opposite signs, the different signs of $\tilde{g}_a$ will lead to different mass arrangements for a specific di-hadron system with different total angular momentum numbers. The $P_{\psi}^N/P^\Lambda_{\psi s}$ states in the scenario 1 have been discussed in detail within the same framework in Ref. \cite{Chen:2022wkh}. However, since both scenarios can not be excluded at present, to give more valuable predictions to the considered heavy flavor meson-baryon and di-hadron systems, in this work, we further add the results of the $P_{\psi}^N$ and $P^\Lambda_{\psi s}$ systems with the inputs from the scenario 2.
\subsection{Flavor-spin symmetry among the $P_{\psi}^N$/$P^\Lambda_{\psi s}$/$H_{\Omega_{ccc}}^N$/$H_{\Omega_{ccc}s}^{\Lambda}$ systems in the single channel formalism}\label{SC}
With the obtained parameters $\tilde{g}_s$ and $\tilde{g}_a$ in both scenarios, we firstly present a single channel calculation to discuss how the flavor-spin symmetry manifests itself among the $P_{\psi}^N$/$P^\Lambda_{\psi s}$/$H_{\Omega_{ccc}}^N$/$H^{\Lambda}_{\Omega_{ccc}s}$ systems.
Here, we list the diagonal matrix elements of the operators $\mathcal{O}^{\text{f}}=\langle\bm{\lambda}_1\cdot\bm{\lambda}_2\rangle$ and $\mathcal{O}^{\text{fs}}=\langle\bm{\lambda}_1\cdot\bm{\lambda}_2\bm{\sigma}_1\cdot\bm{\sigma}_2\rangle$ for the $P_{\psi}^N/P^\Lambda_{\psi s}/H_{\Omega_{ccc}}^N/H_{\Omega_{ccc}s}^{\Lambda}$ systems with different total angular momentum numbers in Table \ref{SCoperators}.
Then the numerical effective potential for a di-hadron $[H_1H_2]_J^I$ system can be directly written as
\begin{eqnarray}
V_{[H_1H_2]_J^I}=\tilde{g}_s\mathcal{O}^{\text{f}}+\tilde{g}_a\mathcal{O}^{\text{fs}}.
\end{eqnarray}
With the numerical effective potentials, we solve the corresponding bound state solutions with the inputs from the scenario 1 and scenario 2, and collect them in Table \ref{SCresults}.

In our convention, the negative and positive effective potentials correspond to the attractive and repulsive forces, respectively. As given in Table \ref{SCoperators}, the $[\Lambda_c\bar{D}]_{\frac{1}{2}}$ , $[\Lambda_c\bar{D}^*]_{\frac{1}{2},\frac{3}{2}}$, $[\Lambda_c\Xi_{cc}]_{0,1}$, and $[\Lambda_c\Xi_{cc}^*]_{1,2}$ systems have repulsive forces and do not have bound state solutions. Besides, the $[\Lambda_c\bar{D}_s]_{\frac{1}{2}}$, $[\Lambda_c\bar{D}_s^*]_{\frac{1}{2},\frac{3}{2}}$, $[\Lambda_c\Omega_{cc}]_{0,1}$, and $[\Lambda_c\Omega_{cc}^*]_{1,2}$ systems have weak attractive forces but not strong enough to form bound states. However, we find that after including the coupled-channel and SU(3) breaking effects, the existences of the $[\Lambda_c\bar{D}_s]_{\frac{1}{2}}$, $[\Lambda_c\bar{D}_s^*]_{\frac{1}{2},\frac{3}{2}}$, $[\Lambda_c\Omega_{cc}]_{0,1}$, and $[\Lambda_c\Omega_{cc}^*]_{1,2}$ bound states are possible, we will discuss this issue in Sec. \ref{Pcs and Hcs}.
\begin{table*}[htbp]
\renewcommand\arraystretch{1.5}
\caption{The diagonal matrix elements of the $\mathcal{O}^{\text{f}}=\langle\bm{\lambda}_1\cdot\bm{\lambda}_2\rangle$ and
$\mathcal{O}^{\text{fs}}=\langle(\bm{\lambda}_1\cdot\bm{\lambda}_2)(\bm{\sigma}_1\cdot\bm{\sigma}_2)\rangle$ for the $P_{\psi}^N/H_{\Omega_{ccc}}^N/P_{\psi s}^{\Lambda}/H^{\Lambda}_{\Omega_{ccc}s}$ systems. We use the superscript A-F on the total angular momentum $J$ to denote six groups of di-hadron systems that have identical effective potentials.}
\begin{tabular}{c|cc|c|ccccccccccc}
\toprule[1pt]
\hline
System&$\mathcal{O}^{\text{f}}$&$\mathcal{O}^{\text{fs}}$&System&$\mathcal{O}^{\text{f}}$&$\mathcal{O}^{\text{fs}}$\\
\hline
$[\Lambda_c\bar{D}]_{\frac{1}{2}}$&$\frac{2}{3}$&0&$[\Lambda_c\Xi_{cc}]_{0,1}$&$\frac{2}{3}$,$\frac{2}{3}$&0,0\\
$[\Lambda_c\bar{D}^*]_{\frac{1}{2},\frac{3}{2}}$&$\frac{2}{3}$,$\frac{2}{3}$&0,0
&$[\Lambda_c\Xi_{cc}^*]_{1,2}$&$\frac{2}{3}$,$\frac{2}{3}$&0,0\\
$[\Sigma_c\bar{D}]_{\frac{1}{2}^{\text{A}}}$&$-\frac{10}{3}$&0
&$[\Sigma_c\Xi_{cc}]_{0^{\text{B}},1}$&$-\frac{10}{3}$,$-\frac{10}{3}$&$-\frac{20}{9}$,$\frac{20}{27}$\\
$[\Sigma_c^*\bar{D}]_{\frac{3}{2}^{\text{A}}}$&$-\frac{10}{3}$&0
&$[\Sigma_c^*\Xi_{cc}]_{1,2^{\text{C}}}$&$-\frac{10}{3}$,$-\frac{10}{3}$&$-\frac{50}{27}$,$\frac{10}{9}$\\
$[\Sigma_c\bar{D}^*]_{\frac{1}{2},\frac{3}{2}^{\text{B}}}$&$-\frac{10}{3}$,$-\frac{10}{3}$&$\frac{40}{9}$,$-\frac{20}{9}$
&$[\Sigma_c\Xi_{cc}^*]_{1,2^{\text{B}}}$&$-\frac{10}{3}$,$-\frac{10}{3}$&$\frac{100}{27}$,$-\frac{20}{9}$\\
$[\Sigma^*_c\bar{D}^*]_{\frac{1}{2}^{\text{D}},\frac{3}{2},\frac{5}{2}^{\text{E}}}$&$-\frac{10}{3}$,$-\frac{10}{3}$,$-\frac{10}{3}$&$\frac{50}{9}$,$\frac{20}{9}$,$-\frac{10}{3}$
&$[\Sigma_c^*\Xi_{cc}^{*}]_{0^{\text{D}},1,2^{\text{C}},3^{\text{E}}}$&$-\frac{10}{3}$,$-\frac{10}{3}$,$-\frac{10}{3}$&$\frac{50}{9}$,$\frac{110}{27}$,$\frac{10}{9}$,$-\frac{10}{3}$\\
\hline
\hline
System&$\mathcal{O}^{\text{f}}$&$\mathcal{O}^{\text{fs}}$&System&$\mathcal{O}^{\text{f}}$&$\mathcal{O}^{\text{fs}}$\\
\hline
$[\Lambda_c\bar{D}_s]_{\frac{1}{2}^{\text{F}}}$&$-\frac{4}{3}$&0
&$[\Lambda_c\Omega_{cc}]_{0^{\text{F}},1^{\text{F}}}$&$-\frac{4}{3}$,$-\frac{4}{3}$&0,0\\
$[\Lambda_c\bar{D}_s^*]_{\frac{1}{2}^{\text{F}},\frac{3}{2}^{\text{F}}}$&$-\frac{4}{3}$,$-\frac{4}{3}$&0,0
&$[\Lambda_c\Omega_{cc}^*]_{1^{\text{F}},2^{\text{F}}}$&$-\frac{4}{3}$,$-\frac{4}{3}$&0,0\\
$[\Xi_c\bar{D}]_{\frac{1}{2}^{\text{A}}}$&$-\frac{10}{3}$&0
&$[\Xi_c\Xi_{cc}]_{0^{\text{A}},1^{\text{A}}}$&$-\frac{10}{3}$,$-\frac{10}{3}$&0,0\\
$[\Xi_c\bar{D}^*]_{\frac{1}{2}^{\text{A}},\frac{3}{2}^{\text{A}}}$&$-\frac{10}{3}$,$-\frac{10}{3}$&0,0
&$[\Xi_c\Xi_{cc}^*]_{1^{\text{A}},2^{\text{A}}}$&$-\frac{10}{3}$,$-\frac{10}{3}$&0,0\\
$[\Xi^{\prime}_c\bar{D}]_{\frac{1}{2}^{\text{A}}}$&$-\frac{10}{3}$&0
&$[\Xi^\prime_c\Xi_{cc}]_{0^{\text{B}},1}$&$-\frac{10}{3}$,$-\frac{10}{3}$&$-\frac{20}{9}$,$\frac{20}{27}$\\
$[\Xi_c^*\bar{D}]_{\frac{3}{2}^{\text{A}}}$&$-\frac{10}{3}$&0
&$[\Xi_c^*\Xi_{cc}]_{1,2^{\text{C}}}$&$-\frac{10}{3}$,$-\frac{10}{3}$&$-\frac{50}{27}$,$\frac{10}{9}$\\
$[\Xi^\prime_c\bar{D}^*]_{\frac{1}{2},\frac{3}{2}^{\text{B}}}$
&$-\frac{10}{3}$,$-\frac{10}{3}$&$\frac{40}{9}$,$-\frac{20}{9}$
&$[\Xi^\prime_c\Xi_{cc}^*]_{1,2^{\text{B}}}$&$-\frac{10}{3}$,$-\frac{10}{3}$&$\frac{100}{27}$,$-\frac{20}{9}$\\
$[\Xi^*_c\bar{D}^*]_{\frac{1}{2}^{\text{D}},\frac{3}{2},\frac{5}{2}^{\text{E}}}$&$-\frac{10}{3}$,$-\frac{10}{3}$,$-\frac{10}{3}$&$\frac{50}{9}$,$\frac{20}{9}$,$-\frac{10}{3}$
&$[\Xi_c^*\Xi_{cc}^{*}]_{0^{\text{D}},1,2^{\text{C}},3^{\text{E}}}$&$-\frac{10}{3}$,$-\frac{10}{3}$,$-\frac{10}{3}$,$-\frac{10}{3}$&$\frac{50}{9}$,$\frac{110}{27}$,$\frac{10}{9}$,$-\frac{10}{3}$\\
\bottomrule[1pt]
\end{tabular}
\label{SCoperators}
\end{table*}
\begin{table*}[htbp]
\renewcommand\arraystretch{1.5}
\caption{The bound state solutions of the $P_{\psi}^N/H_{\Omega_{ccc}}^N/P^\Lambda_{\psi s}/H^{\Lambda}_{\Omega_{ccc}s}$ systems obtained within the single-channel formalism. The results are calculated from the inputs of scenario 1 and scenario 2. We use the superscript A-F on the total angular momentum $J$ to denote six groups of di-hadron systems that share identical effective potentials. All the results are in units of MeV. }
\begin{tabular}{c|cc|c|ccccccccccc}
\toprule[1pt]
&\multicolumn{2}{c|}{Scenario 1}&&\multicolumn{2}{c}{Scenario 2}\\
\hline
System&Mass (MeV)&BE (MeV)&System&Mass (MeV)&BE (MeV)\\
\hline
$[\Sigma_c\bar{D}]_{\frac{1}{2}^{\text{A}}}$&$4312.6$&$-8.1$&$[\Sigma_c\bar{D}]_{\frac{1}{2}^{\text{A}}}$&$4307.6$&$-13.1$\\
$[\Sigma_c^*\bar{D}]_{\frac{3}{2}^{\text{A}}}$&$4376.9$&$-8.5$&$[\Sigma_c^*\bar{D}]_{\frac{3}{2}^{\text{A}}}$&$4371.7$&$-13.6$\\
$[\Sigma_c\bar{D}^*]_{\frac{1}{2},\frac{3}{2}^{\text{B}}}$&$4438.8$,$4457.5$&$-23.2$,$-4.6$&$[\Sigma_c\bar{D}^*]_{\frac{1}{2},\frac{3}{2}^{\text{B}}}$&$4456.8$,$4440.9$&$-5.3$,$-21.1$\\
$[\Sigma^*_c\bar{D}^*]_{\frac{1}{2}^{\text{D}},\frac{3}{2},\frac{5}{2}^{\text{E}}}$&4498.8,4510.3,4523.8&$-27.9$,$-16.4$,$-2.9$
&$[\Sigma^*_c\bar{D}^*]_{\frac{1}{2}^{\text{D}},\frac{3}{2},\frac{5}{2}^{\text{E}}}$&4522.9,4516.6,4501.6&$-3.8$,$-10.1$,$-25.1$\\
\hline
$[\Sigma_c\Xi_{cc}]_{0^{\text{B}},1}$&6060.8,6050.3&$-14.0$,$-24.6$&$[\Sigma_c\Xi_{cc}]_{0^{\text{B}},1}$&6037.5,6048.1&$-37.3$,$-26.8$\\
$[\Sigma_c^*\Xi_{cc}]_{1,2^{\text{C}}}$&6123.7,6112.7&$-15.9$,$-26.8$&$[\Sigma_c^*\Xi_{cc}]_{1,2^{\text{C}}}$&6102.7,6113.2&$-36.9$,$-26.3$\\
$[\Sigma_c\Xi_{cc}^*]_{1,2^{\text{B}}}$&6143.0,6166.0&$-37.5$,$-14.5$&$[\Sigma_c\Xi_{cc}^*]_{1,2^{\text{B}}}$&6162.6,6142.4&$-17.8$,$-38.0$\\
$[\Sigma_c^*\Xi_{cc}^{*}]_{0^{\text{D}},1}$&6198.3,6205.1&$-46.8$,$-40.1$
&$[\Sigma_c^*\Xi_{cc}^{*}]_{0^{\text{D}},1}$&6232.0,6227.7&$-13.2$,$-17.4$\\
$[\Sigma_c^*\Xi_{cc}^{*}]_{2^{\text{C}},3^{\text{E}}}$&6217.7,6233.6&$-27.4$,$-11.6$
&$[\Sigma_c^*\Xi_{cc}^{*}]_{2^{\text{C}},3^{\text{E}}}$&6218.2,6201.9&$-26.9$,$-43.2$\\
\hline
\hline
$[\Xi_c\bar{D}]_{\frac{1}{2}^{\text{A}}}$&4328.1&$-8.2$&$[\Xi_c\bar{D}]_{\frac{1}{2}^{\text{A}}}$&4323.1&$-13.3$\\
$[\Xi_c\bar{D}^*]_{\frac{1}{2}^{\text{A}},\frac{3}{2}^{\text{A}}}$&4468.0,4468.0&$-9.7$,$-9.7$&$[\Xi_c\bar{D}^*]_{\frac{1}{2}^{\text{A}},\frac{3}{2}^{\text{A}}}$&4462.5,4462.5&$-15.1$,$-15.1$\\
$[\Xi^{\prime}_c\bar{D}]_{\frac{1}{2}^{\text{A}}}$&4436.8&$-8.9$&$[\Xi^{\prime}_c\bar{D}]_{\frac{1}{2}^{\text{A}}}$&4431.6&$-14.1$\\
$[\Xi_c^*\bar{D}]_{\frac{3}{2}^{\text{A}}}$&4503.9&$-9.3$&$[\Xi_c^*\bar{D}]_{\frac{3}{2}^{\text{A}}}$&4498.6&$-14.6$\\
$[\Xi^\prime_c\bar{D}^*]_{\frac{1}{2},\frac{3}{2}^{\text{B}}}$&4562.5,4581.8&$-24.5$,$-5.2$&$[\Xi^\prime_c\bar{D}^*]_{\frac{1}{2},\frac{3}{2}^{\text{B}}}$&4581.1,4564.7&$-5.9$,$-22.3$\\
$[\Xi^*_c\bar{D}^*]_{\frac{1}{2}^{\text{D}},\frac{3}{2},\frac{5}{2}^{\text{E}}}$&4625.3,4637.0,4651.2&$-29.2$,$-17.5$,$-3.4$
&$[\Xi^*_c\bar{D}^*]_{\frac{1}{2}^{\text{D}},\frac{3}{2},\frac{5}{2}^{\text{E}}}$&4650.2,4643.6,4628.2&$-4.3$,$-11.0$,$-26.3$\\
\hline
$[\Xi_c\Xi_{cc}]_{0^{\text{A}},1^{\text{A}}}$&6068.5,6068.5&$-22.0$,$-22.0$&$[\Xi_c\Xi_{cc}]_{0^{\text{A}},1^{\text{A}}}$&6061.0,6061.0&$-29.5$,$-29.5$\\
$[\Xi_c\Xi_{cc}^*]_{1^{\text{A}},2^{\text{A}}}$&6173.6,6173.6&$-22.5$,$-22.5$&$[\Xi_c\Xi_{cc}^*]_{1^{\text{A}},2^{\text{A}}}$&6165.9,6165.9&$-30.2$,$-30.2$\\
$[\Xi^\prime_c\Xi_{cc}]_{0^{\text{B}},1}$&6184.7,6173.8&$-15.2$,$-26.1$&$[\Xi^\prime_c\Xi_{cc}]_{0^{\text{B}},1}$&6160.8,6171.6&$-39.1$,$-28.3$\\
$[\Xi_c^*\Xi_{cc}]_{1,2^{\text{C}}}$&6250.3,6239.1&$-17.1$,$-28.3$&$[\Xi_c^*\Xi_{cc}]_{1,2^{\text{C}}}$&6228.8,6239.6&$-38.6$,$-27.8$\\
$[\Xi^\prime_c\Xi_{cc}^*]_{1,2^{\text{B}}}$&6266.2,6289.8&$-39.2$,$-15.7$&$[\Xi^\prime_c\Xi_{cc}^*]_{1,2^{\text{B}}}$&6286.3,6265.7&$-19.1$,$-39.8$\\
$[\Xi_c^*\Xi_{cc}^{*}]_{0^{\text{D}},1}$&6324.3,6331.2&$-48.7$,$-41.8$
&$[\Xi_c^*\Xi_{cc}^{*}]_{0^{\text{D}},1}$&6358.7,6354.3&$-14.3$,$-18.6$\\
$[\Xi_c^*\Xi_{cc}^{*}]_{2^{\text{C}},3^{\text{E}}}$&6344.0,6360.4&$-29.0$,$-12.6$
&$[\Xi_c^*\Xi_{cc}^{*}]_{2^{\text{C}},3^{\text{E}}}$&6344.6,6327.9&$-28.4$,$-45.1$\\
\bottomrule[1pt]
\end{tabular}\label{SCresults}
\end{table*}

From Table \ref{SCoperators}, we also find that the systems
\begin{eqnarray}
\Sigma_c^{(*)}\bar{D}&\leftrightarrow&\Xi_c^{\prime(*)}\bar{D},\\
\Sigma_c^*\bar{D}^*&\leftrightarrow&\Xi_c^{\prime(*)}\bar{D}^*,\\
\Sigma_c^{(*)}\Xi_{cc}&\leftrightarrow&\Xi_c^{\prime(*)}\Xi_{cc},\\
\Sigma_c^{(*)}\Xi_{cc}^*&\leftrightarrow&\Xi_c^{\prime(*)}\Xi_{cc}^*,
\end{eqnarray}
with the same total angular momentum numbers (with the same spin wave functions) have identical contact potentials. This is the main manifestation of the SU(3) symmetry.

The diagonal matrix elements $\mathcal{O}^{\text{fs}}$ for the $[\Xi_c\bar{D}^*]_{\frac{1}{2}}$ and $[\Xi_c\bar{D}^*]_{\frac{3}{2}}$ systems are all 0, the $\Xi_c$ and $\bar{D}^*$ components have vanishing spin-spin interaction from their light degrees of freedom. The inclusion of the spin-spin interaction terms from their heavy degrees of freedom could distinguish the $\Xi_c\bar{D}^*$ molecular states with $J=1/2$ and $J=3/2$, but this is beyond the scope of this work. Thus, if we only consider the interactions introduced from the exchanges of light mesons, the $[\Xi_c\bar{D}^*]_{\frac{1}{2}}$ and $[\Xi_c\bar{D}^*]_{\frac{3}{2}}$ systems have identical effective potential and thus have identical binding energy, as presented in Table \ref{SCresults}.

The degeneracy of a di-hadron system with different total angular momentum numbers also appear in the $[\Lambda_c\bar{D}^*]_{\frac{1}{2},\frac{3}{2}}$, $[\Lambda_c\Xi_{cc}]_{0,1}$, $[\Lambda_c\Xi_{cc}^*]_{1,2}$, $[\Lambda_c\bar{D}_s^*]_{\frac{1}{2},\frac{3}{2}}$, $[\Lambda_c\Omega_{cc}]_{0,1}$, $[\Lambda_c\Omega_{cc}^{*}]_{1,2}$, $[\Xi_c\Xi_{cc}]_{0,1}$, $[\Xi_c\Xi_{cc}^{*}]_{1,2}$ systems, as can be seen from their effective potentials listed in Table \ref{SCoperators}. By checking the results listed in Table \ref{SCresults}, we find that among the above systems, the $[\Xi_c\Xi_{cc}]_{0,1}$ and $[\Xi_c\Xi_{cc}^{*}]_{1,2}$ systems have bound state solutions, the $[\Xi_c\Xi_{cc}]_0$ ($[\Xi_c\Xi_{cc}^*]_1$) and $[\Xi_c\Xi_{cc}]_1$ ($[\Xi_c\Xi_{cc}^*]_2$) have identical binding energy, one needs to introduce the spin-spin interaction terms from their heavy degrees of freedom to discriminate the $[\Xi_c\Xi_{cc}]_0$ ($[\Xi_c\Xi_{cc}^*]_1$) and $[\Xi_c\Xi_{cc}]_1$ ($[\Xi_c\Xi_{cc}^*]_2$) bound states. In this work, we assume that the mass corrections induced from the exchanges of $c\bar{c}$ mesons are very small and neglect this effect.



By simply collecting the systems that have identical attractive effective potentials, we obtain six groups of heavy flavor di-hadron systems, we label these six groups of systems with the superscript A-F on the angular momentum $J$ in Tables \ref{SCoperators} and \ref{SCresults}.

Among the systems in A-F groups, the systems in F group have weak attractive force and can form bound state only if we consider the coupled-channel and SU(3) breaking effects, we will discuss the F group in Sec. \ref{BE discussion}.

As can be seen from Tables \ref{SCoperators} and \ref{SCresults}, on the one hand, in each of the A-E groups, with the same effective potential, the heavy-flavor di-baryon bound state has deeper binding energy than that of the heavy flavor meson-baryon bound state, this result implies the special role of heavy flavor quark components in stabilizing the molecules. On the other hand, since the heavy flavor di-baryon system and the baryon-meson system with identical effective potential in the same group have different binding energies, although they are related via the flavor-spin symmetry, one can not directly identify a heavy flavor di-baryon bound state as the flavor-spin symmetry partner of a heavy flavor baryon-meson bound state through their binding energies, which might be observed in the future. Alternatively, in each group, the flavor-spin symmetry manifests itself very well in the heavy flavor meson-baryon and di-baryon systems, separately. For example, from Table \ref{SCresults}, we further list the binding energies in group A obtained with the inputs from scenario 1 as follows
\begin{eqnarray}
&&[\Xi_c\bar{D}]_{\frac{1}{2}}:-8.2\, \text{MeV}, [\Xi_c\bar{D}^*]_{\frac{1}{2},\frac{3}{2}}:-9.7\, \text{MeV},-9.7\, \text{MeV}\nonumber\\
&&[\Sigma_c\bar{D}]_{\frac{1}{2}}:-8.1\, \text{MeV}, [\Xi^\prime_c\bar{D}]_{\frac{1}{2}}:-8.9\, \text{MeV},\nonumber\\
&&[\Sigma_c^*\bar{D}]_{\frac{3}{2}}:-8.5\, \text{MeV}, [\Xi_c^*\bar{D}]_{\frac{3}{2}}:-9.3\, \text{MeV}, \nonumber
\end{eqnarray}
for the heavy flavor baryon-meson bound states and
\begin{eqnarray}
&&[\Xi_c\Xi_{cc}]_{0,1}:-22.0\,\text{MeV}\,-22.0\,\text{MeV}\,\nonumber\\
&&[\Xi_c\Xi_{cc}^*]_{1,2}:-22.5\,\text{MeV},-22.5\,\text{MeV},\nonumber
\end{eqnarray}
for the heavy flavor di-baryon bound states, respectively. The heavy flavor meson-baryon or di-baryon systems in group A have very similar binding energies, this phenomenon can serve as a fingerprint for the existence of the flavor-spin symmetry. We can also obtain the same conclusion by checking the binding energies of the heavy flavor baryon-meson or di-baryon systems in B-E groups. Besides, this conclusion also applies to the results in A-E groups obtained with the inputs from scenario 2, as can be checked from Table \ref{SCresults}.

From our single-channel calculation, the flavor-spin symmetry manifests itself very well in the $P_{\psi}^N/H_{\Omega_{ccc}}^N/P^\Lambda_{\psi s}/H^{\Lambda}_{\Omega_{ccc}s}$ systems, however, different from the $P_{\psi}^N$ and $H_{\Omega_{ccc}}^N$ systems, the inclusion of strange quark will violate the SU(3) flavor symmetry, this might introduce significant differences to the mass spectra of the $P^\Lambda_{\psi s}$ and $H^{\Lambda}_{\Omega_{ccc}s}$ systems. Besides, the coupled-channel effect may also shift the masses of the bound states obtained from the single-channel calculation in the $P_{\psi}^N/H_{\Omega_{ccc}}^N/P^\Lambda_{\psi s}/H^{\Lambda}_{\Omega_{ccc}s}$ systems. In the following, we discuss the influences of these two effects on the violations of flavor-spin symmetry.

\section{$P_{\psi}^N/P_{\psi}^\Lambda/H_{\Omega_{ccc}}^N/H^{\Lambda}_{\Omega_{ccc}s}$ spectra in the coupled-channel formalism}\label{MC}
The flavors of light quark components in the $P_{\psi}^N$ and $H_{\Omega_{ccc}}^N$ systems are the same, the replacement of a $u/d$ quark within the $P_{\psi}^N$ and $H_{\Omega_{ccc}^N}$ systems to an $s$ quark lead to the $P^\Lambda_{\psi s}$ and $H^{\Lambda}_{\Omega_{ccc}s}$ systems. In this section, we firstly discuss the molecular spectra of the $P_{\psi}^N$ and $H_{\Omega_{ccc}}^N$ systems by including the coupled-channel effect. Then we proceed to discuss the $P^\Lambda_{\psi s}$ and $H^{\Lambda}_{\Omega_{ccc}s}$ systems. To give more complete descriptions to the spectra of the $P^\Lambda_{\psi s}$ and $H^{\Lambda}_{\Omega_{ccc}s}$ systems, we will consider both the coupled-channel effect and SU(3) breaking effect.
\subsection{Spectra of the $P_{\psi}^N$ and $H_{\Omega_{ccc}}^N$ systems}\label{Pc and Hc}
\begin{table*}[htbp]
\renewcommand\arraystretch{1.5}
\caption{The matrix elements of $[\langle\bm{\lambda}_1\cdot\bm{\lambda}_2\rangle,\langle\bm{\lambda}_1\cdot\bm{\lambda}_2\bm{\sigma}_1\cdot\bm{\sigma}_2\rangle]$ for the di-baryon channels associated with the effective potential matrices $\mathbb{V}_0^{H_{\Omega_{ccc}}^N}$ and $\mathbb{V}_2^{H_{\Omega_{ccc}}^N}$.}
\begin{tabular}{cccc|ccccccccccc}
\toprule[1pt]
\multicolumn{4}{c|}{$\mathbb{V}_0^{H_{\Omega_{ccc}}^N}$}&\multicolumn{5}{c}{$\mathbb{V}_2^{H_{\Omega_{ccc}}^N}$}\\
\hline
Channel&$\Lambda_c\Xi_{cc}$&$\Sigma_c\Xi_{cc}$&$\Sigma_c^*\Xi_{cc}^*$
&Channel&$\Lambda_c\Xi_{cc}^*$&$\Sigma_c^*\Xi_{cc}$&$\Sigma_c\Xi_{cc}^*$&$\Sigma_c^*\Xi_{cc}^*$\\
\hline
$\Lambda_c\Xi_{cc}$      &$[\frac{2}{3},0]$      &$[0,1]$      &$[0,2\sqrt{2}]$
&$\Lambda_c\Xi_{cc}^*$&$[\frac{2}{3},0]$&$[0,-4]$&$[0,2]$&$[0,4]$\\
$\Sigma_c\Xi_{cc}$&&$[-\frac{10}{3},-\frac{20}{9}]$&$[0,\frac{20\sqrt{2}}{9}]$
&$\Sigma_c^*\Xi_{cc}$&&$[-\frac{10}{3},\frac{10}{9}]$&$[0,-\frac{20}{9}]$&$[0,-\frac{40}{9}]$\\
$\Sigma_c^*\Xi_{cc}^*$&&&$[-\frac{10}{3},\frac{50}{9}]$
&$\Sigma_c\Xi_{cc}^*$&&&$[-\frac{10}{3},-\frac{20}{9}]$&$[0,\frac{20}{9}]$\\
&&&&$\Sigma_c^*\Xi_{cc}^*$&&&&$[-\frac{10}{3},\frac{10}{9}]$\\
\hline
\bottomrule[1pt]
\end{tabular}\label{VPC02}
\end{table*}
\begin{table*}[htbp]
\renewcommand\arraystretch{1.5}
\caption{The matrix elements of $[\langle\bm{\lambda}_1\cdot\bm{\lambda}_2\rangle,\langle\bm{\lambda}_1\cdot\bm{\lambda}_2\bm{\sigma}_1\cdot\bm{\sigma}_2\rangle]$ for the di-baryon channels associated with the effective potential matrix $\mathbb{V}_1^{H_{\Omega_{ccc}}^N}$.}
\begin{tabular}{ccccccccccccccc}
\toprule[1pt]
\multicolumn{7}{c}{$\mathbb{V}_1^{H_{\Omega_{ccc}}^N}$}\\
\hline
Channel&$\Lambda_c\Xi_{cc}$&$\Lambda_c\Xi_{cc}^*$&$\Sigma_c\Xi_{cc}$&$\Sigma_c^*\Xi_{cc}$&$\Sigma_c\Xi_{cc}^*$&$\Sigma_c^*\Xi_{cc}^*$\\
\hline
$\Lambda_c\Xi_{cc}$&$[\frac{2}{3},0]$&$[0,0]$&$[0,-\frac{1}{3}]$&$[0,-\frac{2\sqrt{2}}{3}]$&$[0,\frac{4\sqrt{2}}{3}]$&$[0,\frac{2\sqrt{10}}{3}]$\\
$\Lambda_c\Xi_{cc}^*$&&$[\frac{2}{3},0]$&$[0,\frac{8\sqrt{2}}{3}]$&$[0,-\frac{4}{3}]$&$[0,-\frac{10}{3}]$&$[0,\frac{4\sqrt{5}}{3}]$\\
$\Sigma_c\Xi_{cc}$&&&$[-\frac{10}{3},\frac{20}{27}]$&$[0,-\frac{20\sqrt{2}}{27}]$&$[0,-\frac{80\sqrt{2}}{27}]$&$[0,\frac{20\sqrt{10}}{27}]$\\
$\Sigma_c^*\Xi_{cc}$&&&&$[-\frac{10}{3},-\frac{50}{27}]$&$[0,-\frac{20}{27}]$&$[0,-\frac{40\sqrt{5}}{27}]$\\
$\Sigma_c\Xi_{cc}^*$&&&&&$[-\frac{10}{3},\frac{100}{27}]$&$[0,\frac{20\sqrt{5}}{27}$]\\
$\Sigma_c^*\Xi_{cc}^*$&&&&&&$[-\frac{10}{3},\frac{110}{27}]$\\
\bottomrule[1pt]
\end{tabular}\label{VPC1}
\end{table*}


With the same framework, we have calculated the effective potential matrices ($\mathbb{V}_{\frac{1}{2}}^{P_{\psi}^N}$, $\mathbb{V}^{P_{\psi}^N}_{\frac{3}{2}}$) in Table II of Ref. \cite{Chen:2022wkh}, we refer the interested readers to Ref. \cite{Chen:2022wkh} for more details. We collect the numerical results of the effective potential matrices $\mathbb{V}_{0}^{H_{\Omega_{ccc}}^N}$, $\mathbb{V}_2^{H_{\Omega_{ccc}}^N}$, and $\mathbb{V}_1^{H_{\Omega_{ccc}}^N}$ for the $H_{\Omega_{ccc}}^N$ systems in Tables \ref{VPC02} and \ref{VPC1}. As can be seen from Tables \ref{VPC02} and \ref{VPC1}, the diagonal matrix elements of matrices $\mathbb{V}_0^{H_{\Omega_{ccc}}^N}$, $\mathbb{V}_1^{H_{\Omega_{ccc}}^N}$, and $\mathbb{V}_2^{H_{\Omega_{ccc}}^N}$ all have non-vanishing central terms, and some of them have corrections from the spin-spin interaction terms. On the contrary, the central terms of the off-diagonal matrix elements in $\mathbb{V}_{0}^{H_{\Omega_{ccc}}^N}$, $\mathbb{V}_2^{H_{\Omega_{ccc}}^N}$, and $\mathbb{V}_1^{H_{\Omega_{ccc}}^N}$ are all 0, the different $H_{\Omega_{ccc}}^N$ systems with the same total angular momentum $J$ couple to each other through the spin-spin interaction terms. The above conclusions also apply to the ($\mathbb{V}_{\frac{1}{2}}^{P_{\psi}^N}$, $\mathbb{V}^{P_{\psi}^N}_{\frac{3}{2}}$) matrices as can be checked from Table II of Ref. \cite{Chen:2022wkh}. Due to the small coupling parameter $\tilde{g}_a$, we may anticipate that the mass corrections of the $P_{\psi}^N/H_{\Omega_{ccc}}^N$ bound states induced from the coupled-channel effect would be relatively small.

\begin{figure*}[!htbp]
    \centering
    \includegraphics[width=1.0\linewidth]{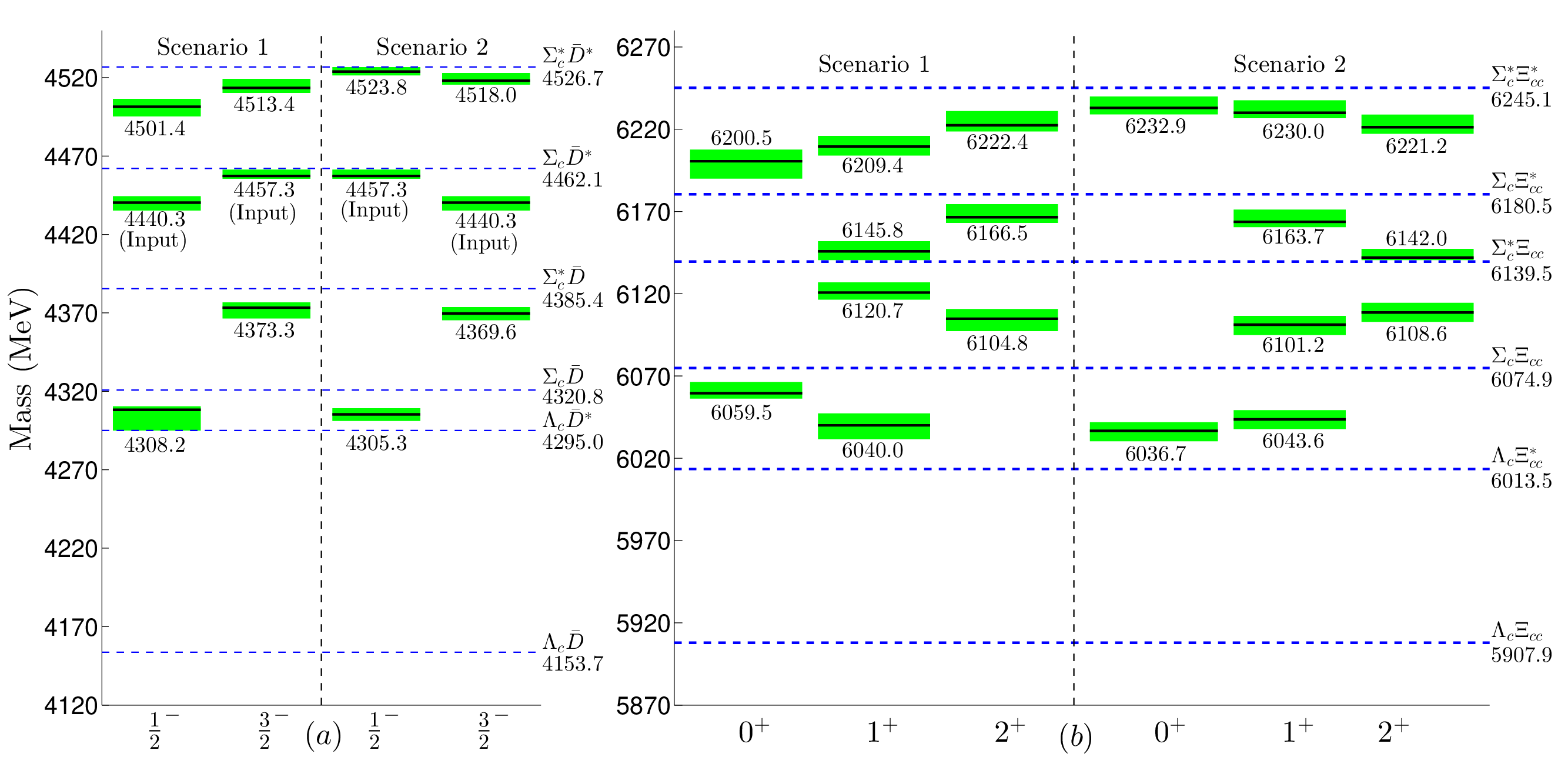}
    \caption{The mass spectra of the $P_{\psi}^N$ and $H_{\Omega_{ccc}}^N$ systems obtained from our coupled-channel formalism. The results of the $P_{\psi}^N$ bound states for the scenario 1 and scenario 2 are illustrated in the left and right sides of (a), respectively. The results of the $H_{\Omega_{ccc}}^N$ bound states for the scenario 1 and scenario 2 are illustrated in the left and right sides of (b), respectively. We plot the central values of the $P_{\psi}^N$ and $H_{\Omega_{ccc}}^N$ bound states with black lines and label the corresponding numerical values. The theoretical errors are illustrated with small green bands, they are introduced by considering the experimental errors \cite{LHCb:2019kea} from the masses and widths of the $P_{\psi}^N(4440)$ and $P_{\psi}^N(4457)$ states.}
    \label{PcHc}
\end{figure*}

We plot the bound state solutions of the $P_{\psi}^N$ and $H_{\Omega_{ccc}}^N$ systems from our coupled-channel calculations in Fig. \ref{PcHc}. The results of the $P_{\psi}^N$ bound states for the scenario 1 and scenario 2 are illustrated in the left and right sides of Fig. \ref{PcHc} (a), respectively. The results of the $H_{\Omega_{ccc}}^N$ bound states for the scenario 1 and scenario 2 are illustrated in the left and right sides of Fig. \ref{PcHc} (b), respectively. We illustrate the central values of the masses of the $P_{\psi}^N/H_{\Omega_{ccc}}^N$ bound states with black lines, and their numerical values are also given in Fig. \ref{PcHc}. We consider the experimental errors from the masses and widths \cite{LHCb:2019kea} of the $P_{\psi}^N(4440)$ and $P_{\psi}^N(4457)$ states to estimate the theoretical errors of the predicted $P_{\psi}^N$/$H_{\Omega_{ccc}}^N$ states, and plot them with small green bands in Fig. \ref{PcHc}.

For the states $[\Sigma_c^*\bar{D}]_{\frac{5}{2}}$ and $[\Sigma_c^*\Xi_{cc}^*]_{3}$ with the highest total angular momentum numbers, we do not need to consider the coupled-channel effect, their results have already been calculated in our single-channel formalism and listed in Table \ref{SCresults}, we do not further illustrate them in Fig. \ref{PcHc}.

As given in Eqs. (\ref{scenario 1}-\ref{scenario 2}), the coupling parameter $\tilde{g}_s$ in the central term solved from the scenario 1 is very close to that of the scenario 2. Besides, in both scenarios, the values of $\tilde{g}_s$ are much larger than that of the $\tilde{g}_a$, thus, the central terms dominant the total effective potentials of all the $P_{\psi}^N$ and $H_{\Omega_{ccc}}^N$ systems. The similar and large $\tilde{g}_s$ values in both scenarios are the reasons that in these two scenarios, the numbers of the obtained bound states in the $P_{\psi}^N$ or $H_{\Omega_{ccc}}^N$ mass spectrum are the same.

The central term is also related to the flavor matrix element $\mathcal{O}^{\text{f}}$, which is determined by the flavor structure of the light quark components in each of the $P_{\psi}^N$ or $H_{\Omega_{ccc}}^N$ system. Thus, the $P_{\psi}^N$ and $H_{\Omega_{ccc}}^N$ states with the same total isospin will have identical central term. Due to the dominant role of central terms and the larger reduced masses in the $H_{\Omega_{ccc}}^N$ systems, we can say that the existence of $P_{\psi}^N$ bound state with isospin $I$ also implies the existence of $H_{\Omega_{ccc}}^N$ bound states with the same isospin.


The mass spectra of the $P_{\psi}^N$ and $H_{\Omega_{ccc}}^N$ states constructed from Eq. (\ref{potential1}) can be described by the following picture. The central term provide the dominant attractive force to bind the $\Sigma_c^{(*)}$ baryon with $\bar{D}^{(*)}$ meson or $\Xi_{cc}^{(*)}$ baryon, while the spin-spin interaction term will further shift the mass of the obtained bound state by several or a few tens of MeV.

Since the values of $\tilde{g}_a$ determined from scenarios 1 and 2 have opposite signs, the arrangements of the $P_{\psi}^N$ and $H_{\Omega_{ccc}}^N$ spectra are different. As illustrated in Fig. \ref{PcHc}, in scenario 1, the masses of the bound states that are mainly composed of the $\Sigma^{(*)}_c\bar{D}^*$ system increase as the total angular momentum $J$ increases. Correspondingly, in this scenario, the masses of the bound states that are mainly composed of the $\Sigma^{(*)}_c\Xi_c$ system decrease as the $J$ increases, while the masses of the bound states that are mainly composed of the $\Sigma^{(*)}_c\Xi_{cc}^*$ system increase as the $J$ increases.

On the contrary, as illustrated in the right sides of Figs. \ref{PcHc} (a) and (b), in the scenario 2, the tendencies discussed in scenario 1 are all reversed.

We also notice that the mass spectra of the $P_{\psi}^N$ and $H_{\Omega_{ccc}}^N$ systems have also been discussed in Ref. \cite{Pan:2019skd} based on a effective field theory that respects heavy quark symmetry. They use a contact-range Lagrangian to describe the interactions in the heavy flavor meson-baryon and di-baryon systems. In their framework, the $\Sigma_c^{(*)}$, $\bar{D}^{(*)}$, and $\Xi_{cc}^{(*)}$ hadrons are described by introducing the corresponding superfields. By comparing the effective potentials of the $P_{\psi}^N/H_{\Omega_{ccc}}^N$ systems in this work and the effective potentials of the $P_{\psi}^N/H_{\Omega_{ccc}}^N$ systems in Ref. \cite{Pan:2019skd}, we find that the effective potentials of the $P_{\psi}^N/H_{\Omega_{ccc}}^N$ systems obtained in these two works are consistent with each other. This is due to the fact that we still describe the interactions in the heavy flavor meson-baryon and di-baryon systems at hadron level, but we reexpress their effective potentials in terms of quark-level language. Correspondingly, the mass arrangements of the $P_{\psi}^N$ and $H_{\Omega_{ccc}}^N$ mass spectra in these two works are also consistent with each other.

The arrangements of the $P_{\psi}^N$ and $H_{\Omega_{ccc}}^N$ spectra would be the signatures of the flavor-spin symmetry among the interactions in the $P_{\psi}^N$ and $H_{\Omega_{ccc}}^N$ systems, thus, the confirmation of these arrangements would give strong support to not only the molecular nature of the observed $P_{\psi}^N$ states, but also a general flavor-spin symmetry among the interactions of the $P_{\psi}^N$ and $H_{\Omega_{ccc}}^N$ states.

\subsection{Spectra of the $P_{\psi s}^{\Lambda}$ and $H^{\Lambda}_{\Omega_{ccc}s}$ systems}\label{Pcs and Hcs}
\begin{table*}[htbp]
\renewcommand\arraystretch{1.5}
\caption{The matrix elements of $[\langle\bm{\lambda}_1\cdot\bm{\lambda}_2\rangle,\langle\bm{\lambda}_1\cdot\bm{\lambda}_2\bm{\sigma}_1\cdot\bm{\sigma}_2\rangle]$ for the di-baryon channels associated with the effective potential matrices $\mathbb{V}_0^{H^{\Lambda}_{\Omega_{ccc}s}}$, $\mathbb{V}_2^{H^{\Lambda}_{\Omega_{ccc}s}}$, $\mathbb{V}_1^{H^{\Lambda}_{\Omega_{ccc}s}}$, and $\mathbb{V}_1^{\prime H^{\Lambda}_{\Omega_{ccc}s}}$.}
\begin{tabular}{c|cccc|c|ccccccccc}
\toprule[1pt]
\multicolumn{5}{c}{$\mathbb{V}_0^{H_{\Omega_{ccc}s}^{\Lambda}}$}&\multicolumn{6}{c}{$\mathbb{V}_2^{H_{\Omega_{ccc}s}^{\Lambda}}$}\\
\hline
Channel&$\Lambda_c\Omega_{cc}$&$\Xi_c\Xi_{cc}$&$\Xi_c^\prime\Xi_{cc}$&$\Xi_c^*\Xi_{cc}^*$
&Channel&$\Lambda_c\Omega_{cc}^*$&$\Xi_c\Xi_{cc}^*$&$\Xi_c^*\Xi_{cc}$&$\Xi_c^\prime\Xi_{cc}^*$&$\Xi_c^*\Xi_{cc}^*$\\
$\Lambda_c\Omega_{cc}$&$[-\frac{4}{3},0]$&$[2\sqrt{2},0]$&$[0,\sqrt{\frac{2}{3}}]$&$[0,\frac{4}{\sqrt{3}}]$
&$\Lambda_c\Omega_{cc}^*$&$[-\frac{4}{3},0]$&$[2\sqrt{2},0]$&$[0,-4\sqrt{\frac{2}{3}}]$&$[0,2\sqrt{\frac{2}{3}}]$&$[0,4\sqrt{\frac{2}{3}}]$\\
$\Xi_c\Xi_{cc}$&&$[-\frac{10}{3},0]$&$[0,\frac{1}{\sqrt{3}}]$&$[0,2\sqrt{\frac{2}{3}}]$
&$\Xi_c\Xi_{cc}^*$&&$[-\frac{10}{3},0]$&$[0,-\frac{4}{\sqrt{3}}]$&$[0,\frac{2}{\sqrt{3}}]$&$[0,\frac{4}{\sqrt{3}}]$\\
$\Xi_c^\prime\Xi_{cc}$&&&$[-\frac{10}{3},-\frac{20}{9}]$&$[0,\frac{20\sqrt{2}}{9}]$
&$\Xi_c^*\Xi_{cc}$&&&$[-\frac{10}{3},\frac{10}{9}]$&$[0,-\frac{20}{9}]$&$[0,-\frac{40}{9}]$\\
$\Xi_c^*\Xi_{cc}^*$&&&&$[-\frac{10}{3},\frac{50}{9}]$
&$\Xi_c^\prime\Xi_{cc}^*$&&&&$[-\frac{10}{3},-\frac{20}{9}]$&$[0,\frac{20}{9}]$\\
&&&&&$\Xi_c^*\Xi_{cc}^*$&&&&&$[-\frac{10}{3},\frac{10}{9}]$\\
\hline
\multicolumn{5}{c}{$\mathbb{V}_1^{H_{\Omega_{ccc}s}^{\Lambda}}$}&\multicolumn{5}{c}{$\mathbb{V}_1^{\prime H_{\Omega_{ccc}s}^{\Lambda}}$}\\
\hline
Channel&$\Lambda_c\Omega_{cc}$&$\Lambda_c\Omega_{cc}^*$&$\Xi_c\Xi_{cc}$&$\Xi_c\Xi_{cc}^*$
&Channel&$\Xi_c^\prime\Xi_{cc}$&$\Xi_c^*\Xi_{cc}$&$\Xi_c^\prime\Xi_{cc}^*$&$\Xi_c^*\Xi_{cc}^*$\\
$\Lambda_c\Omega_{cc}$&$[-\frac{4}{3},0]$&$[0,0]$&$[2\sqrt{2},0]$&$[0,0]$
&$\Xi_c^\prime\Xi_{cc}$&$[-\frac{10}{3},\frac{20}{27}]$&$[0,-\frac{20\sqrt{2}}{27}]$&$[0,-\frac{80\sqrt{2}}{27}]$&$[0,\frac{20\sqrt{10}}{27}]$\\
$\Lambda_c\Omega_{cc}^*$&&$[-\frac{4}{3},0]$&$[0,0]$&$[2\sqrt{2},0]$
&$\Xi_c^*\Xi_{cc}$&&$[-\frac{10}{3},-\frac{50}{27}]$&$[0,-\frac{20}{27}]$&$[0,-\frac{40\sqrt{5}}{27}]$\\
$\Xi_c\Xi_{cc}$&&&$[-\frac{10}{3},0]$&$[0,0]$
&$\Xi_c^\prime\Xi_{cc}^*$&&&$[-\frac{10}{3},\frac{100}{27}]$&$[0,\frac{20\sqrt{5}}{27}]$\\
$\Xi_c\Xi_{cc}^*$&&&&$[-\frac{10}{3},0]$
&$\Xi_c^*\Xi_{cc}^*$&&&&$[-\frac{10}{3},\frac{110}{27}]$\\
\bottomrule[1pt]
\end{tabular}\label{VHcs}
\end{table*}
The effective potential matrices $(\mathbb{V}_{\frac{1}{2}}^{P^\Lambda_{\psi s}},\mathbb{V}_{\frac{3}{2}}^{P^\Lambda_{\psi s}})$ of the $P^\Lambda_{\psi s}$ systems have been given in Table III of Ref. \cite{Chen:2022wkh}, we refer the interested readers to Ref. \cite{Chen:2022wkh} for more details.

The effective potential matrices ($\mathbb{V}_0^{H^{\Lambda}_{\Omega_{ccc}s}}$,$\mathbb{V}_1^{H^{\Lambda}_{\Omega_{ccc}s}}$,$\mathbb{V}_1^{\prime H^{\Lambda}_{\Omega_{ccc}s}}$,$\mathbb{V}_2^{H^{\Lambda}_{\Omega_{ccc}s}}$) of the $H^{\Lambda}_{\Omega_{ccc}s}$ systems are presented in Table \ref{VHcs}. Here, as given in Table \ref{coupled-channels}, to study the $J=1$ systems, we need to consider the interactions coupled from eight channels. Finding the pole positions of the $J=1$ $H^{\Lambda}_{\Omega_{ccc}s}$ states from this eight coupled-channel calculation is relatively time consuming and unpractical (in our calculation, it takes about ten times longer to find the bound state solutions from an eight coupled-channel LSE than that of a seven coupled-channel LSE). Alternatively, from Table \ref{systems}, one can find that the thresholds from the lowest $\Lambda_c\Omega_{cc}$ channel to the highest $\Xi_c^*\Xi_{cc}^*$ channel range from 6064.5 to 6373.0 MeV, the threshold gap is more than 300 MeV. In general, only the channels with their thresholds close to the considered channel would give significant corrections to the interaction of the considered channel, besides, from our calculation, we find that the $(\Lambda_c\Omega_{cc},\Lambda_c\Omega_{cc}^*,\Xi_c\Xi_{cc},\Xi_c\Xi_{cc}^*)$ channels couple to the $(\Xi_c^\prime\Xi_{cc},\Xi_c^*\Xi_{cc},\Xi_c^\prime\Xi_{cc}^*,\Xi_c^*\Xi_{cc}^*)$ channels only through the spin-spin interaction terms, due to the small spin-spin coupling parameter $\tilde{g}_a$, we can expect that the corrections from the couplings of the lower four channels with the higher four channels would be small. Thus, we simplify the calculation of the $J=1$ $H^{\Lambda}_{\Omega_{ccc}s}$ system by dividing the considered eight channels in Table \ref{coupled-channels} into two groups, i.e., the $(\Lambda_c\Omega_{cc},\Lambda_c\Omega_{cc}^*,\Xi_c\Xi_{cc},\Xi_c\Xi_{cc}^*)$ and $(\Xi_c^\prime\Xi_{cc},\Xi_c^*\Xi_{cc},\Xi_c^\prime\Xi_{cc}^*,\Xi_c^*\Xi_{cc}^*)$. We present the corresponding effective potential matrices $\mathbb{V}_{1}^{H^{\Lambda}_{\Omega_{ccc}s}}$ and $\mathbb{V}_1^{\prime H^{\Lambda}_{\Omega_{ccc}s}}$ in Table \ref{VHcs}.

There exists an important difference between the effective potential matrices in the $H_{\Omega_{ccc}}^N$ and $H^{\Lambda}_{\Omega_{ccc}s}$ systems. As given in Table \ref{VHcs}, for the off-diagonal matrix elements, the effective potentials of the $\Lambda_c\Omega_{cc}^{(*)}-\Xi_c\Xi_{cc}^{(*)}$ channel in the $H_{\Omega_{ccc}s}^{\Lambda}$ systems with $J=0$, 1, and 2 consist of non-vanishing central terms. Due to the large coupling parameter $\tilde{g}_s$, the coupling from these channels will give considerable corrections to the mass spectra of the $H_{\Omega_{ccc}s}^{\Lambda}$ systems with $J=0$, 1, and 2. As discussed in Ref. \cite{Chen:2022wkh}, similar coupling also exists in the $P_{\psi s}^{\Lambda}$ systems, i.e., the $\Lambda_c\bar{D}_s^{(*)}-\Xi_c\bar{D}^{(*)}$ coupling with $J=1/2$ or $3/2$. 

When checking the diagonal matrix elements listed in Tables II and III of Ref. \cite{Chen:2022wkh} for the $P_{\psi}^N$ and $P^\Lambda_{\psi s}$ systems and the diagonal matrix elements listed in Tables \ref{VPC02}, \ref{VPC1}, and \ref{VHcs} for the $H_{\Omega_{ccc}}^N$ and $H^{\Lambda}_{\Omega_{ccc}s}$ systems, we find that their dominant components are from the exchanges of the non-strange light scalar meson currents, i.e, from the matrix elements
$\langle\lambda_1^i\lambda_2^i\rangle$, with $i$ sums from $1$ to $3$.
Since the interactions of the off-diagonal channel $\Lambda_c\bar{D}_s^{(*)}-\Xi_c\bar{D}^{(*)}$ and $\Lambda_c\Omega_{cc}^{(*)}-\Xi_{c}\Xi_{cc}^{(*)}$ are introduced from the exchanges of the strange scalar meson currents, i.e., from the matrix elements
$\langle\lambda_1^j\lambda_2^j\rangle$, with $j$ sums from $4$ to $7$. Comparing with the exchanges of the non-strange light meson currents, the off-diagonal matrix elements should be suppressed by the masses of the exchanged strange mesons. Here, we quantify the SU(3) breaking effect by multiplying a $g_x$ factor on all the off-diagonal matrix elements that describing the effective potentials of the $\Lambda_c\bar{D}_s^{(*)}-\Xi_c\bar{D}^{(*)}$ and $\Lambda_c\Omega_{cc}^{(*)}-\Xi_{c}\Xi_{cc}^{(*)}$ couplings (one can refer to Ref. \cite{Chen:2022wkh} for more details). We assume that $g_x$ is in the range $[0,1]$, at $g_x=0$, the $\Lambda_c\bar{D}_s^{(*)}$ ($\Lambda_c\Omega_{cc}^{(*)}$) channel does not couple to the $\Xi_c\bar{D}^{(*)}$ ($\Xi_c\Xi_{cc}^{(*)}$) channel. While at $g_x=1$, the coupling strength between the $\Lambda_c\bar{D}_s^{(*)}$ ($\Lambda_c\Omega_{cc}^{(*)}$) channel and $\Xi_c\bar{D}^{(*)}$ ($\Xi_c\Xi_{cc}^{(*)}$) channel is in its SU(3) limit.

From our previous multi-channel study on the $P_{\psi s}^{\Lambda}$ systems \cite{Chen:2022wkh}, we find that the coupled-channel effect only have considerable corrections to the bound state solutions that are related to the $\Lambda_c\bar{D}_s^{(*)}$ and $\Xi_{c}\bar{D}^{(*)}$ channels due to their considerable coupling induced from the central terms. Consequently, the bound state solutions will have significant dependences on the SU(3) breaking factor $g_x$. But for the rest of channels that can only couple to the other channels through the spin-spin interaction terms, the coupled-channel effect will give small corrections to their bound state solutions, and have very tiny dependences on the parameter $g_x$.

The above conclusions also applies to the $H_{\Omega_{ccc}s}^{\Lambda}$ systems as well, from our multi-channel calculations on the $H_{\Omega_{ccc}s}^{\Lambda}$ systems with $J=0$, 1, and 2, we find that only the bound state solutions that are related to the $\Lambda_c\Omega_{cc}^{(*)}$ and $\Xi_c\Xi_{cc}^{(*)}$ channels have significant dependences on the SU(3) breaking factor $g_x$. Thus, we firstly discuss the special roles of the $\Lambda_c\bar{D}_s^{(*)}-\Xi_c\bar{D}^{(*)}$ and $\Lambda_c\Omega_{cc}^{(*)}-\Xi_c\Xi_{cc}^{(*)}$ couplings and their $g_x$-dependences, then we present the full mass spectra of $P^\Lambda_{\psi s}$ and $H^{\Lambda}_{\Omega_{ccc}s}$ systems.
\begin{figure*}[!htbp]
    \centering
    \includegraphics[width=1.0\linewidth]{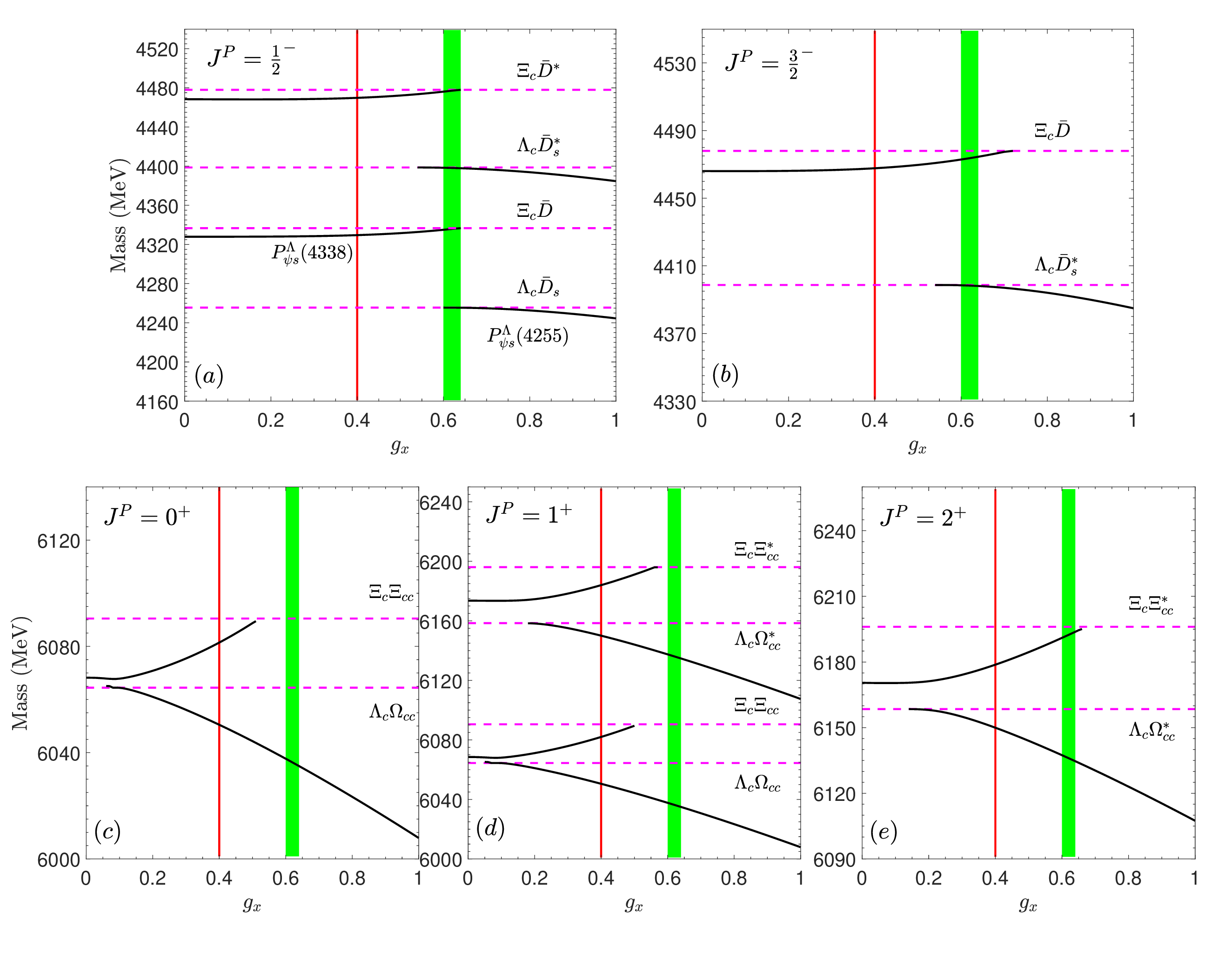}
    \caption{With the inputs from scenario 1, as the $g_x$ increases from 0 to 1, the variations of the masses for the $P_{\psi s}^{\Lambda}$ bound states that are attributed to the $\Lambda_c\bar{D}_s^{(*)}$ and $\Xi_c\bar{D}^{(*)}$ channels with $J^P=\frac{1}{2}^-$ and $\frac{3}{2}^-$ are plotted in (a) and (b), respectively. The variations of the masses for the $H_{\Omega_{ccc}s}^{\Lambda}$ bound states that are attributed to the $\Lambda_c\Omega_{cc}^{(*)}$ and $\Xi_c\Xi_{cc}^{(*)}$ channels with $J^P=0^+$, $1^+$, and $2^+$ are plotted in (c), (d), and (e), respectively. We plot the green bands with $g_x$ at [0.60,0.62] to denote the region that the $P_{\psi s}^{\Lambda}(4338)$ and $P_{\psi s}^{\Lambda}(4255)$ can coexist, and we plot the red lines with $g_x=0.40$ to denote the value that only the $P_{\psi s}^{\Lambda}(4338)$ exists.}
    \label{case1}
\end{figure*}

\begin{figure*}[!htbp]
    \centering
    \includegraphics[width=1.0\linewidth]{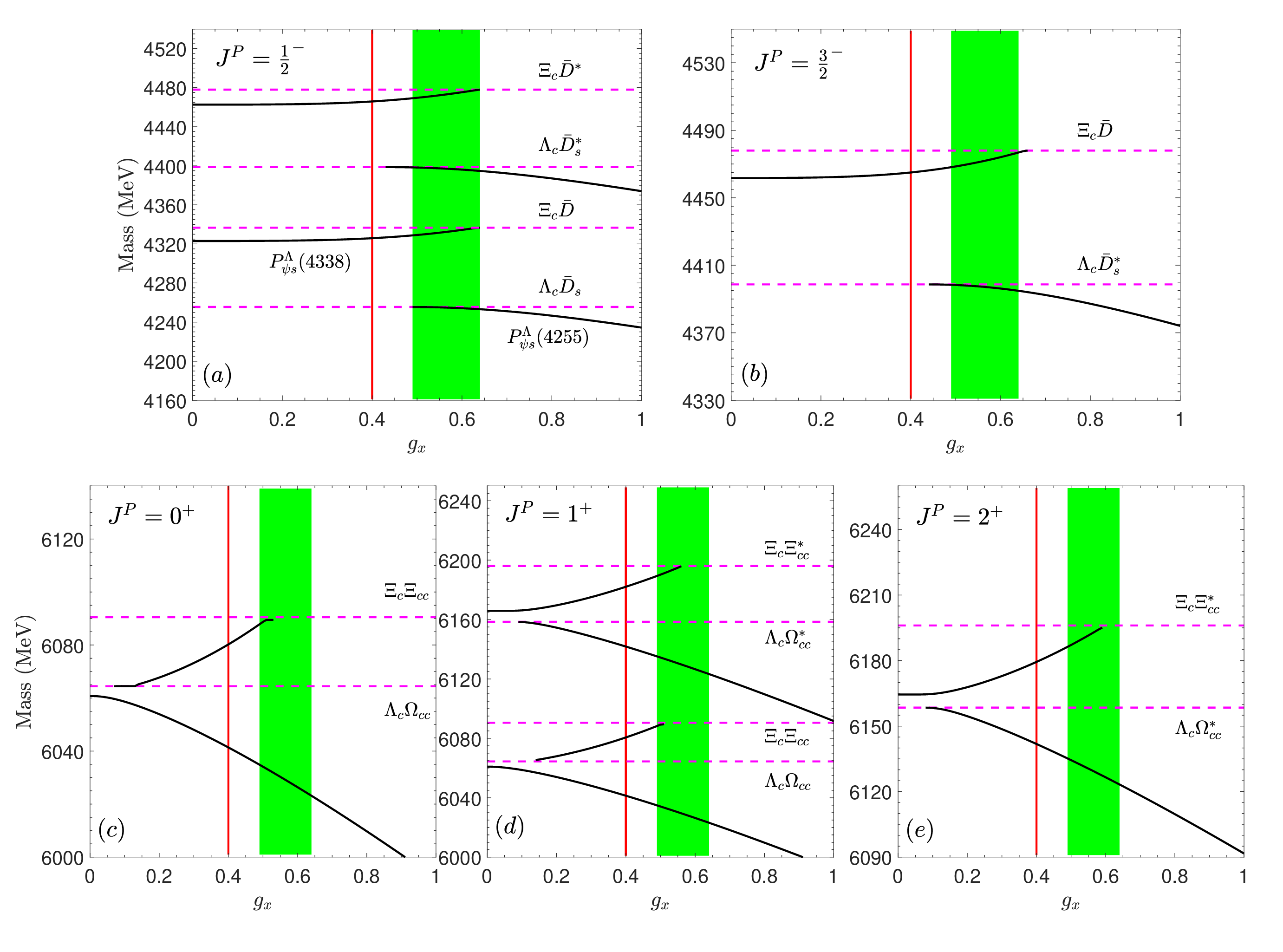}
    \caption{With the inputs from scenario 2, as the $g_x$ increases from 0 to 1, the variations of the masses for the $P_{\psi s}^\Lambda$ bound states that are attributed to the $\Lambda_c\bar{D}_s^{(*)}$ and $\Xi_c\bar{D}^{(*)}$ channels with $J^P=\frac{1}{2}^-$ and $\frac{3}{2}^-$ are plotted in (a) and (b), respectively. The variations of the masses for the $H^{\Lambda}_{\Omega_{ccc}s}$ bound states that are attributed to the $\Lambda_c\Omega_{cc}^{(*)}$ and $\Xi_c\Xi_{cc}^{(*)}$ channels with $J^P=0^+$, $1^+$, and $2^+$ are plotted in (c), (d), and (e), respectively. We plot the green bands with $g_x$ at [0.60,0.62] to denote the region that the $P_{\psi s}^{\Lambda}(4338)$ and $P_{\psi s}^{\Lambda}(4255)$ can coexist, and we plot the red lines with $g_x=0.40$ to denote the value that only the $P_{\psi s}^{\Lambda}(4338)$ exists.}
    \label{case2}
\end{figure*}
With the inputs from scenario 1, we run the $g_x$ value in the range $[0,1]$, we perform our multi-channel calculations and present the $g_x$-dependences of the masses for the bound states that are attributed to the $\Lambda_c\bar{D}_s^{(*)}$ and $\Xi_c\bar{D}^{(*)}$ channels for the $P^\Lambda_{\psi s}$ systems with $J^P=\frac{1}{2}^-$ and $\frac{3}{2}^-$ in Figs. \ref{case1} (a) and (b), respectively. Similarly, we present the $g_x$-dependences of the masses for the bound states that are attributed to the $\Lambda_c\Omega_{cc}^{(*)}$ and $\Xi_c\Xi_{cc}^{(*)}$ channels for the $H^{\Lambda}_{\Omega_{ccc}s}$ systems with $J^{P}=0^+$, $1^{+}$, and $2^{+}$ in Figs. \ref{case1} (c), (d), and (e), respectively. The results with the inputs from scenario 2 are also illustrated in Fig. \ref{case2}. In the following, we mainly discuss the results obtained from scenario 1 in Fig. \ref{case1}, the results obtained from scenario 2 in Fig. \ref{case2} can be discussed in a similar way.

In Ref. \cite{LHCb:2022ogu}, the LHCb collaboration reported the $P^\Lambda_{\psi s}(4338)$, this state could be the $\Xi_c\bar{D}$ molecular state. Besides, from the $J/\Psi\Lambda$ invariant spectrum, there might be a $P^\Lambda_{\psi s}(4255)$ structure near the $\Lambda_c\bar{D}_s$ threshold, further confirmation on this state is still needed. In Fig. \ref{case1} (a), we use the green band with $g_x$ at [0.60,0.62] to label the region that the $P^\Lambda_{\psi s}(4338)$ and $P^\Lambda_{\psi s}(4225)$ states can coexist, then we select the $g_x$ value at 0.40 to label the results that only the $P^\Lambda_{\psi s}(4338)$ exists, while the $P^\Lambda_{\psi s}$ does not exist. The $g_x=0.40$ is plotted in Fig. \ref{case1} with red lines. Then we use the same $g_x$ region and value to label the $J^P=\frac{3}{2}^-$ $P^\Lambda_{\psi s}$ system.

Then we proceed to use the same $g_x$ region and value obtained from the $J^P=\frac{1}{2}^-$ $P^\Lambda_{\psi s}$ system to label the results of the $H_{\Omega_{ccc}s}^{\Lambda}$ systems with $J^P=0^+$, $1^+$, and $2^+$ in Figs. \ref{case1} (c), (d), and (e), respectively. As plotted in Figs. \ref{case1} (c), (d), and (e), for the $H_{\Omega_{ccc} s}^{\Lambda}$ systems, at $g_x=0.40$, the bound states that are attributed to the $\Lambda_c\Omega_{cc}^{(*)}$ channel with $J=0$, 1, 2 and the bound states that are attributed to the $\Xi_{c}\Xi_{cc}^{(*)}$ with $J=0$, 1, 2 can coexist. But when the $g_x$ is at $[0.60,0.62]$, for the $J^P=0^+$ and $1^+$ $H_{\Omega_{ccc} s}^{\Lambda}$ systems, only the bound states that are attributed to the $\Lambda_c\Omega_{cc}^{(*)}$ channel exist, the bound states that are attributed to the $\Xi_c\Xi_{cc}^{(*)}$ disappear. But for the $J^P=2^+$ system, the bound state that is attributed to the $\Lambda_c\Omega_{cc}^*$ channel and the bound state that is attributed to the $\Xi_c\Xi_{cc}^*$ can coexist.

From Fig. \ref{case1}, we also find that as the $g_x$ increases, the masses of the bound states in the $H_{\Omega_{ccc}s}^{\Lambda}$ systems increase or decrease more rapidly than that of the bound states in the $P^\Lambda_{\psi s}$ systems. We can understand our results from two aspects. On the one hand, the flavor of the light quark components for the $\Lambda_c\bar{D}_s^{(*)}$ and $\Xi_c\bar{D}^{(*)}$ systems are identical to that of the $\Lambda_c\Omega_{cc}^{(*)}$ and $\Xi_c\Xi_{cc}^{(*)}$ systems, respectively. From Table III of Ref. \cite{Chen:2022wkh} and Table \ref{VHcs}, we can see that the matrix elements of the $\Lambda_c\bar{D}^{(*}_s-\Lambda_c\bar{D}_s^{(*)}$, $\Xi_{c}\bar{D}^{(*)}-\Xi_c\bar{D}^{(*)}$, and $\Lambda_c\bar{D}^{(*)}_s-\Xi_c\bar{D}_s^{(*)}$ in the $P^\Lambda_{\psi s}$ systems are identical to the matrix elements of the $\Lambda_c\Omega_{cc}^{(*)}-\Lambda_c\Omega_{cc}^{(*)}$, $\Xi_c\Xi_{cc}^{(*)}-\Xi_c\Xi_{cc}^{(*)}$, and $\Lambda_c\Omega_{cc}^{(*)}-\Xi_c\Xi_{cc}^{(*)}$ in the $H^{\Lambda}_{\Omega_{ccc}s}$ systems, respectively.
These matrix elements only consist of the central terms, and their spin-spin interaction terms vanish. The two coupled-channel system ($\Lambda_c\bar{D}_s^{(*)}$, $\Xi_{c}\bar{D}^{(*)}$) and ($\Lambda_c\Omega_{cc}^{(*)}$, $\Xi_c\Xi_{cc}^{(*)}$) with different $J$ have identical effective potentials. However, the $\Lambda_c\Omega_{cc}^{(*)}$ and $\Xi_c\Xi_{cc}^{(*)}$ systems have larger reduced masses than that of the $\Lambda_c\bar{D}_s^{(*)}$ and $\Xi_c\bar{D}^{(*)}$ systems, respectively. Thus, if the attractive forces in the ($\Lambda_c\bar{D}_s^{(*)}$, $\Xi_{c}\bar{D}^{(*)}$) and ($\Lambda_c\Omega_{cc}^{(*)}$, $\Xi_c\Xi_{cc}^{(*)}$) systems increase (decrease) in the same way, the absolute values of the binding energies for the systems with heavier reduced masses will increase (decrease) more significantly.

On the other hand, as can be found from Table \ref{systems}, the mass gaps of the $\Xi_c\bar{D}-\Lambda_c\bar{D}_s$ ($\Xi_c\bar{D}^{*}-\Lambda_c\bar{D}_s^*$) and $\Xi_c\Xi_{cc}-\Lambda_c\Omega_{cc}$ ($\Lambda_c\Omega_{cc}^{*}-\Xi_c\Xi_{cc}^*$) are 81.2 (79.3) and 26.0 (37.6) MeV, respectively. The $\Lambda_c\Omega_{cc}^{(*)}$ and $\Xi_c\Xi_{cc}^{(*)}$ channels lie much closer to each other than that of the $\Lambda_c\bar{D}_s^{(*)}$ and $\Xi_c\bar{D}^{(*)}$ channels. Thus, with the same effective potentials, the $\Lambda_c\Omega_{cc}^{(*)}-\Xi_c\Xi_{cc}^{(*)}$ couplings in the $H^{\Lambda}_{\Omega_{ccc}s}$ systems are more stronger than that of the $\Lambda_c\bar{D}^{(*)}-\Xi_c\bar{D}^{(*)}$ couplings in the $P^\Lambda_{\psi s}$ systems.

\begin{figure*}[!htbp]
    \centering
    \includegraphics[width=1.0\linewidth]{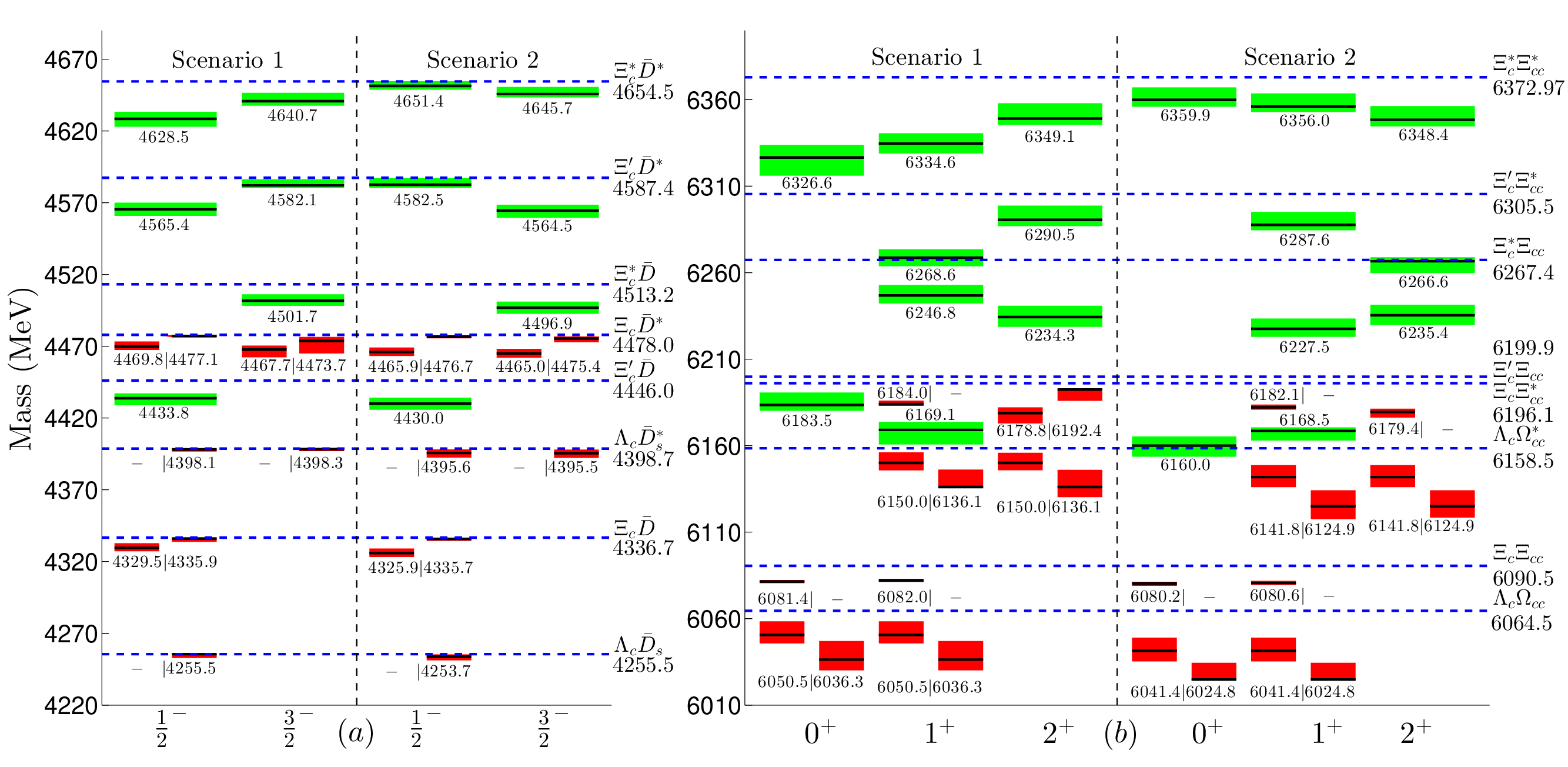}
    \caption{The mass spectra of the $P^\Lambda_{\psi s}$ and $H^{\Lambda}_{\Omega_{ccc}s}$ systems obtained from our coupled-channel formalism. The results of the $P^\Lambda_{\psi s}$ bound states for the scenario 1 and scenario 2 are illustrated in the left and right sides of (a), respectively. The results of the $H^{\Lambda}_{\Omega_{ccc}s}$ bound states for the scenario 1 and scenario 2 are illustrated in the left and right sides of (b), respectively. We plot the central values of the $P^\Lambda_{\psi s}$ and $H^{\Lambda}_{\Omega_{ccc}s}$ bound states with black lines and label the corresponding numerical values. The theoretical errors are introduced by considering the experimental errors from the masses and widths of the $P_{\psi}^N(4440)$ and $P_{\psi}^N(4457)$. The theoretical errors are labeled with green bands. The bound states that are mainly attributed to the $\Lambda_c\bar{D}^{(*)}_s-\Xi_c\bar{D}^{(*)}$ and $\Lambda_c\Omega_{cc}^{(*)}-\Xi_c\Xi_{cc}^{(*)}$ couplings are calculated at $g_x=0.40$ (left) and $g_x=0.62$ (right), we use the ``$-$'' to denote that the bound state do not exist at this $g_x$ value, the theoretical errors of these bound states are labeled with red bands.}
    \label{PcsHcs}
\end{figure*}
We present the complete $P^\Lambda_{\psi s}$ and $H^{\Lambda}_{\Omega_{ccc}s}$ mass spectra from our multi-channel calculations in Fig. \ref{PcsHcs}. Similarly, the bound states $[\Xi_c^*\bar{D}^*]_{\frac{5}{2}}$ and $[\Xi_c^*\Xi_{cc}^*]_3$ with the highest total angular momentum are calculated in the single-channel formalism, their results have been given in Table \ref{SCresults}, we do not further illustrate them in Fig. \ref{PcsHcs}.

Two sets of scenarios given in Eqs. (\ref{scenario 1}-\ref{scenario 2}) are plotted in the left and right sides of Fig. \ref{PcsHcs} (a) and Fig. \ref{PcsHcs} (b). We illustrate the central values of the masses for the $P_{\psi s}^{\Lambda}$ and $H^{\Lambda}_{\Omega_{ccc}s}$ bound states with black lines, and also present their numerical values in Fig. \ref{PcsHcs}. Here, the bound states that are mainly attributed to the $\Lambda_c\bar{D}_s^{(*)}-\Xi_c\bar{D}^{(*)}$ and $\Lambda_c\Omega_{cc}^{(*)}-\Xi_c\Xi_{cc}^{(*)}$ couplings are calculated at two $g_x$ values. As illustrated in Fig. \ref{PcsHcs}, the left and right values are calculated at $g_x=0.40$ and $g_x=0.62$, respectively. We use the ``-'' to denote that the bound state do not exist at this $g_x$ value. The theoretical errors are estimated by considering the experimental errors of the masses and widths \cite{LHCb:2019kea} of the $P_{\psi}^N(4440)$ and $P_{\psi}^N(4457)$ states. We use the red bands to label the theoretical errors of the bound states that are mainly attributed to the $\Lambda_c\bar{D}_s^{(*)}$, $\Xi_c\bar{D}^{(*)}$, $\Lambda_c\Omega_{cc}^{(*)}$, and $\Xi_c\Xi_{cc}^{(*)}$ interactions, while the theoretical errors for the rest of the bound states are still labelled with green bands.

With the parameters in scenario 1 and scenario 2, we also obtain two types of the mass spectra for the $P_{\psi s}^{\Lambda}$ and $H_{\Omega_{ccc}s}^{\Lambda}$ systems. The mass arrangements of the bound states in both $P^{\Lambda}_{\psi s}$ systems and $H_{\Omega_{ccc}s}^{\Lambda}$ systems would be crucial tests to the existence of the flavor-spin symmetry. Specifically, if one can simultaneously observe the increase of the masses as the function of $J$ in the $\Xi_c^\prime\bar{D}^*$ ($\frac{1}{2}^-$, $\frac{3}{2}^-$) multiplet and the decrease of the masses as the function of $J$ in the $\Xi_c^\prime\Xi_{cc}$ ($0^+$, $1^+$) multiplet, then such an observation can definitely serve as a fingerprint of their molecular configurations, and also give strong support to the flavor-spin symmetry among the $P_{\psi s}^{\Lambda}$ and $H^{\Lambda}_{\Omega_{ccc}s}$ molecule community.

In Figs. \ref{PcsHcs} (a) and (b), if we subtract the bound states labeled with red error bars and only focus on the bound states labeled with green error bars, by comparing the mass arrangements of these $P^\Lambda_{\psi s}$ and $H^{\Lambda}_{\Omega_{ccc}s}$ bound states with that of the $P_{\psi}^N$ and $H_{\Omega_{ccc}}^N$ bound states in Fig. \ref{PcHc}, we find that in each scenario, the relative positions of $P_{\psi}^N$ ($H_{\Omega_{ccc}}^N$) bound states in the $P_{\psi}^N$ ($H_{\Omega_{ccc}}^N$) mass spectrum are very similar to the relative positions of $P^\Lambda_{\psi s}$ ($H^{\Lambda}_{\Omega_{ccc}s}$) bound states in the $P^\Lambda_{\psi s}$ ($H^{\Lambda}_{\Omega_{ccc}s}$) mass spectrum, this is exactly the manifestation of flavor-spin symmetry.
\subsection{Binding energies as another fingerprint of the flavor-spin symmetry}\label{BE discussion}
\begin{table*}[htbp]
\renewcommand\arraystretch{1.5}
\caption{The central values of binding energies for the systems in groups A-F (labeled in Table \ref{SCoperators} and \ref{SCresults}) calculated from single-channel (SC) formalism and multi-channel (MC) formalism in the scenario 1 and scenario 2. The bound states that are mainly attributed to the $\Lambda_c\bar{D}_s^{(*)}-\Xi_{c}\bar{D}^{(*)}$ and $\Lambda_c\Omega_{cc}^{(*)}-\Xi_c\Xi_{cc}^{(*)}$ couplings are calculated at $g_x=0.40$ (upper row) and $g_x=0.62$ (lower row), we use the ``-'' to denote that the bound state do not exist at this $g_x$ value. We use the superscript ``$\dagger$'' to label the states that have good flavor-spin symmetry after including the coupled-channel and SU(3) breaking effects. All the values are in units of MeV.}
\begin{tabular}{ccccccccccccccc}
\toprule[1pt]
&\multicolumn{2}{c}{Scenario 1}&\multicolumn{2}{c}{Scenario 2}&&\multicolumn{2}{c}{Scenario 1}&\multicolumn{2}{c}{Scenario 2}\\
System&BE (SC)&BE (MC)&BE (SC)&BE (MC)&System&BE (SC)&BE (MC)&BE (SC)&BE (MC)\\
\hline
$[\Sigma_c\bar{D}]^{\dagger}_{\frac{1}{2}^{\text{A}}}$     &$-8.1$&$-12.6$&$-13.1$&$-15.5$
&$[\Sigma^*_c\bar{D}]^{\dagger}_{\frac{3}{2}^\text{A}}$  &$-8.5$&$-12.6$&$-13.6$&$-15.8$\\
\multirow{2}{*}{$[\Xi_c\bar{D}]_{\frac{1}{2}^\text{A}}$}        &\multirow{2}{*}{$-8.2$}&$-7.1$&\multirow{2}{*}{$-13.1$}&$-10.8$
&\multirow{2}{*}{$[\Xi_c\bar{D}^*]_{\frac{1}{2}^\text{A}}$}       &$\multirow{2}{*}{-9.7}$&$-8.2$&$\multirow{2}{*}{-15.1}$&$-12.1$ \\
&&$-0.8$&&$-1.0$  &&&$-0.9$&&$-1.3$\\
\multirow{2}{*}{$[\Xi_c\bar{D}^*]_{\frac{3}{2}^\text{A}}$}       &\multirow{2}{*}{$-9.7$}&$-10.3$&\multirow{2}{*}{$-15.1$}&$-13.0$
&\multirow{2}{*}{$[\Xi_c^\prime\bar{D}]^{\dagger}_{\frac{1}{2}^\text{A}}$} &\multirow{2}{*}{$-8.9$}&\multirow{2}{*}{$-12.2$}&\multirow{2}{*}{$-14.1$}&\multirow{2}{*}{$-16.2$}\\
&&$-4.3$&&$-2.6$ &&&&&\\
$[\Xi_c^*\bar{D}]_{\frac{3}{2}^\text{A}}$     &$-9.3$&$-11.0$&$-14.6$&$-16.2$&\\
\multirow{2}{*}{$[\Xi_c\Xi_{cc}]_{0^{\text{A}}}$}        &\multirow{2}{*}{$-22.0$}&$-9.0$&\multirow{2}{*}{$-29.5$}&$-10.2$
&\multirow{2}{*}{$[\Xi_c\Xi_{cc}]_{1^{\text{A}}}$}       &\multirow{2}{*}{$-22.0$}&$-8.5$&\multirow{2}{*}{$-29.5$}&$-9.9$\\
&&$-$&&$-$  &&&$-$&&$-$\\
\multirow{2}{*}{$[\Xi_c\Xi_{cc}^*]_{1^{\text{A}}}$}      &\multirow{2}{*}{$-22.5$}&$-12.1$&\multirow{2}{*}{$-30.2$}&$-14.0$
&\multirow{2}{*}{$[\Xi_c\Xi_{cc}^*]_{2^{\text{A}}}$}     &\multirow{2}{*}{$-22.5$}&$-17.3$&\multirow{2}{*}{$-30.2$}&$-16.7$\\
&&$-$&&$-$ &&&$-3.7$&&$-$\\
\hline
$[\Sigma_c\bar{D}^*]^{\dagger}_{\frac{3}{2}^\text{B}}$   &$-4.6$&$-4.8$&$-21.1$&$-21.8$
&$[\Xi_c^\prime\bar{D}^*]^{\dagger}_{\frac{3}{2}^\text{B}}$  &$-5.2$&$-5.2$&$-22.3$&$-22.9$\\
$[\Sigma_c\Xi_{cc}]_{0^\text{B}}^{\dagger}$           &$-14.0$&$-15.3$&$-37.3$&$-38.2$
&$[\Sigma_c\Xi_{cc}^*]_{2^{\text{B}}}^{\dagger}$      &$-14.5$&$-14.0$&$-38.0$&$-38.5$\\
$[\Xi_{c}^\prime\Xi_{cc}]_{0^\text{B}}^{\dagger}$   &$-15.2$&$-16.4$&$-39.1$&$-39.9$
&$[\Xi_c^\prime\Xi^*_{cc}]_{2^{\text{B}}}^{\dagger}$  &$-15.7$&$-15.0$&$-39.8$&$-39.6$\\
\hline
$[\Sigma_c^*\Xi_{cc}]_{2^{\text{C}}}$      &$-26.8$&$-34.8$&$-26.3$&$-30.9$
&$[\Sigma_c^*\Xi_{cc}^*]_{2^{\text{C}}}$     &$-27.4$&$-22.7$&$-26.9$&$-24.0$\\
$[\Xi_c^*\Xi_{cc}]_{2^{\text{C}}}$         &$-28.3$&$-31.8$&$-27.8$&$-31.7$
&$[\Xi_c^*\Xi_{cc}^*]_{2^{\text{C}}}$        &$-29.0$&$-23.8$&$-28.4$&$-24.4$\\
\hline
$[\Sigma_c^*\bar{D}^*]_{\frac{1}{2}^{\text{D}}}^{\dagger}$  &$-27.9$&$-25.3$&$-3.8$&$-2.8$
&$[\Xi_c^*\bar{D}^*]_{\frac{1}{2}^\text{D}}^{\dagger}$        &$-29.2$&$-26.0$&$-4.3$&$-3.2$\\
$[\Sigma_c^*\Xi_{cc}^*]_{0^{\text{D}}}^{\dagger}$           &$-46.8$&$-44.7$&$-13.2$&$-12.2$
&$[\Xi_c^*\Xi_{cc}^*]_{0^{\text{D}}}^{\dagger}$               &$-48.7$&$-46.4$&$-14.3$&$-13.2$\\
\hline
$[\Sigma_c^*\bar{D}^*]_{\frac{5}{2}^{\text{E}}}^{\dagger}$  &$-2.9$&$-2.9$&$-25.1$&$-25.1$
&$[\Xi_c^*\bar{D}^*]_{\frac{5}{2}^{\text{E}}}^{\dagger}$      &$-3.4$&$-3.4$&$26.3$&$-26.3$\\
$[\Sigma_c^*\Xi_{cc}^*]_{3^{\text{E}}}^{\dagger}$            &$-11.6$&$-11.6$&$-43.2$&$-43.2$
&$[\Xi_c^*\Xi_{cc}^*]_{3^\text{E}}^{\dagger}$                &$-12.6$&$-12.6$&$-45.1$&$-45.1$\\
\hline
\multirow{2}{*}{$[\Lambda_c\bar{D}_s]_{\frac{1}{2}^{\text{F}}}$}   &$-$&$-$&$-$&$-$
&\multirow{2}{*}{$[\Lambda_c\bar{D}_s^{*}]_{\frac{1}{2}^{\text{F}}}$} &$-$&$-$&$-$&$-$\\
&$-$&$-0.0$&$-$&$-1.7$   &&$-$&$-0.5$&$-$&$-3.1$&\\
\multirow{2}{*}{$[\Lambda_c\bar{D}_s^{*}]_{\frac{3}{2}^{\text{F}}}$}
&$-$&$-$&$-$&$-$ &&&&&\\
&$-$&$-0.3$&$-$&$-3.2$&  &&&&&\\
\multirow{2}{*}{$[\Lambda_c\Omega_{cc}]_{0^{\text{F}}}$}               &$-$&$-14.0$&$-$&$-23.1$
&\multirow{2}{*}{$[\Lambda_c\Omega_{cc}]_{1^{\text{F}}}$}
&$-$&$-14.0$&$-$&$-23.1$\\
&$-$&$-28.1$&$-$&$-39.7$  &&$-$&$-28.1$&$-$&$-39.7$\\
\multirow{2}{*}{$[\Lambda_c\Omega_{cc}^*]_{1^{\text{F}}}$}             &$-$&$-8.4$&$-$&$-16.6$
&\multirow{2}{*}{$[\Lambda_c\Omega_{cc}^*]_{2^{\text{F}}}$}            &$-$&$-8.5$&$-$&$-16.6$\\
&$-$&$-22.3$&$-$&$-33.5$  &&$-$&$-22.4$&$-$&$-33.5$\\
\bottomrule[1pt]
\end{tabular}\label{BE}
\end{table*}
Apart from testing the arrangements of the mass spectra in the $P_{\psi}^N/P^\Lambda_{\psi s}/H_{\Omega_{ccc}}^N/H^{\Lambda}_{\Omega_{ccc}s}$ systems, checking the binding energies of the heavy flavor meson-baryon or di-baryon systems that are attributed to the identical effective potentials is another way to test the flavor-spin symmetry.

In Table \ref{SCoperators} and \ref{SCresults}, we label the heavy flavor meson-baryon and di-baryon systems that have identical effective potentials with superscript A-F on the total angular momentum $J$. We collect the binding energies of the systems in groups A-F in Table \ref{BE}. The results are obtained in the scenario 1 and scenario 2 within the single-channel formalism (Table \ref{SCresults}) and multi-channel formalism (Fig. \ref{PcHc} and Fig. \ref{PcsHcs}). The results that are mainly attributed to the $\Lambda_c\bar{D}^{(*)}_s$, $\Xi_c\bar{D}^{(*)}$, $\Lambda_c\Omega_{cc}^{(*)}$, and $\Xi_c\Xi_{cc}^{(*)}$ interactions significantly depend on the SU(3) breaking factor $g_x$, we present their results with $g_x=0.40$ (upper row) and $g_x=0.62$ (lower row) in Table \ref{BE}.

In the following, we mainly discuss the results obtained with the inputs from scenario 1, the results obtained with the inputs from scenario 2 can be discussed in a similar way.

For a heavy flavor meson-baryon bound state and a di-baryon bound state with identical effective potential in the single channel case, since the reduced mass of the di-baryon system is heavier than that of the meson-baryon system, the binding of the di-baryon system is more deeper than that of the meson-baryon system, so the binding energies of the di-baryon system is different from that of the meson-baryon system. But for two different heavy flavor meson-baryon or di-baryon bound states with identical effective potential in the single channel case, since they have comparable reduced masses, they will have very similar binding energies. This is the manifestation of the flavor-spin symmetry. Thus, for the systems of the A-F groups listed in Table \ref{BE}, we will compare the binding energies of the meson-baryon and di-baryon systems separately.

From the single channel results collected in Table \ref{BE}, we find that the flavor-spin symmetry manifests itself very well in the A-F groups for the heavy flavor meson-baryon and di-baryon systems, separately. In each group, the binding energies of the heavy flavor meson-baryon systems are very close to each other, similarly, the binding energies of the heavy flavor di-baryon systems are very close to each other, too. For example, in the single channel formalism, the central values of the binding energies for the $[\Sigma_c\bar{D}]_{\frac{1}{2}}$ ($P_{\psi}^N(4312)$) and $[\Sigma_c^*\bar{D}]_{\frac{3}{2}}$ ($P_{\psi}^N(4380)$) \cite{LHCb:2015yax,LHCb:2019kea} bound states are $-8.1$ and $-8.5$ MeV, respectively, very close to each other. Experimentally, the central values of these two systems are $-8.9$ and $-6.2$ MeV, respectively. This consistence can serve as an important evidence of the flavor-spin symmetry. Although the binding energies of the $[\Sigma_c\bar{D}]_{\frac{1}{2}}$ and $[\Sigma_c^*\bar{D}]_{\frac{3}{2}}$ systems are $-12.6$ and $-12.6$ MeV in scenario 2, respectively, and slightly deviated from the experimental values, but we should emphasis that the exact values of our calculation of course will depend on the form factor we adopted in our model, but the closeness of the binding energies between the $[\Sigma_c\bar{D}]_{\frac{1}{2}}$ and $[\Sigma_c^*\bar{D}]_{\frac{3}{2}}$ systems is still maintained. This conclusion is model independent and is the key point of the flavor-spin symmetry.

However, the inclusion of the coupled-channel effect will violate the flavor-spin symmetry in some groups. Specifically, as shown in Table \ref{BE}, from our multi-channel calculation, the systems in groups B, D, and E maintain flavor-spin symmetry, i.e., in each of the three groups, the binding energies of the meson-baryon systems are very close to each other, and the binding energies of the di-baryon systems are very close to each other, too. But for the systems in groups A, C, F, the flavor-symmetry is violated (from the results in B, D, and E groups, we roughly say that the flavor-spin symmetry is violated if the difference of the binding energies between two meson-baryon systems or two di-baryon systems is more than 3.0 MeV).

As shown in Table \ref{BE}, from group A, the binding energy of the $[\Sigma_c\bar{D}]_{\frac{1}{2}}$ system is $-12.6$ MeV in the multi-channel calculation. However, for the $[\Xi_c\bar{D}]_{\frac{1}{2}}$ system, after the inclusion of the coupled-channel and SU(3) breaking effect, its binding energy is obtained as $-7.1$ and $-0.8$ MeV with $g_x=0.40$ and $g_x=0.62$, respectively. The difference of the binding energies between the $[\Sigma_c\bar{D}]_{\frac{1}{2}}$ and $[\Xi_c\bar{D}]_{\frac{1}{2}}$ is larger than 3.0 MeV, and thus the flavor-spin symmetry is violated. Similar violation also appear in the systems of group F. The SU(3) breaking effect is the primary violation source of the flavor-spin symmetry in the systems of groups A and F.


Without the SU(3) breaking effect, the coupled-channel effect itself can also lead to considerable flavor-spin symmetry violation. For the systems in group C, as presented in Table \ref{BE}, the binding energies of the $[\Sigma_c^*\Xi_{cc}]_2$, $[\Sigma_c\Xi_{cc}^*]_2$, $[\Xi_c^*\Xi_{cc}]_{2}$, and $[\Xi_c^*\Xi_{cc}^*]_2$ are $-26.8$, $-27.4$, $-28.3$, and $-29.0$ MeV respectively in the single-channel formalism, comparable with each other. But in the multi-channel calculation, the binding energies of the $[\Sigma_c^*\Xi_{cc}]_2$ and $[\Xi_c^*\Xi_{cc}]_2$ systems become $-34.8$ and $-31.8$ MeV, respectively, while the binding energies of the $[\Sigma_c^*\Xi_{cc}]^*_2$ and $[\Xi_c^*\Xi_{cc}^*]_2$ systems are $-22.7$ and $-23.8$ MeV, respectively. The flavor-spin symmetry violation from the coupled-channel effect can reach up to 10 MeV. However, for the systems in groups B, D, and E, the flavor-spin symmetry is still maintained even if we include the coupled-channel effect. Thus, we conclude that the violation of the flavor-spin symmetry from the coupled-channel effect depends on the specific systems. The systems in groups B, D, E, the $[\Sigma_c\bar{D}]_{\frac{1}{2}}$ and $[\Sigma_c^*\bar{D}]_{\frac{3}{2}}$ systems in group A, we label these systems with superscript ``$\dagger$'', they are ideal bound state candidates for testing the flavor-spin symmetry among the $P_{\psi}^N/H_{\Omega_{ccc}}^N/P_{\psi s}^{\Lambda}/H^{\Lambda}_{\Omega_{ccc}s}$ molecular community.

\section{Summary}\label{summary}
In this work, based on a contact lagrangian possessing the SU(3) flavor symmetry and SU(2) spin symmetry, we discuss the flavor-spin symmetry of the interactions among the $P_{\psi}^N/P^\Lambda_{\psi s}/H_{\Omega_{ccc}}^N/H^{\Lambda}_{\Omega_{ccc}s}$ systems within a unified framework. The flavors of the light quark components for the pentaquark systems $P_{\psi}^N$ and $P^\Lambda_{\psi s}$ are identical to that of the hexaquark systems $H_{\Omega_{ccc}}^N$ and $H^{\Lambda}_{\Omega_{ccc}s}$, respectively. The interactions of these systems are expected to share very similar interactions. This work is devoted to clarify the similarities and differences of the interactions among the $P_{\psi}^N/P^\Lambda_{\psi s}/H_{\Omega_{ccc}}^N/H^{\Lambda}_{\Omega_{ccc}s}$ systems.

The parameters $\tilde{g}_s$ and $\tilde{g}_a$ are determined in two scenarios, i.e., the $J^P$ numbers of the $P_{\psi}^N(4440)$ and $P_{\psi}^N(4457)$ are $\frac{1}{2}^-$ and $\frac{3}{2}$ respectively in scenario 1, and the $J^P$ numbers of the $P_{\psi}^N(4440)$ and $P_{\psi}^N(4457)$ are $\frac{3}{2}^-$ and $\frac{1}{2}^-$ respectively in scenario 2. The obtained values of the $\tilde{g}_s$ are much larger than the values of the $\tilde{g}_a$, thus, the central term dominant the formation of the bound states.

We firstly perform a single channel calculation to obtain the effective potentials and mass spectra of the $P_{\psi}^N/P^\Lambda_{\psi s}/H_{\Omega_{ccc}}^N/H^{\Lambda}_{\Omega_{ccc}s}$ systems.  In the single-channel formalism, since the matrix elements $\mathcal{O}^{\text{f}}$ in the central terms only depend on the flavor wave functions of the $P_{\psi}^N/P^\Lambda_{\psi s}/H_{\Omega_{ccc}}^N/H^{\Lambda}_{\Omega_{ccc}s}$ systems, the $P_{\psi}^N$ ($P^\Lambda_{\psi s}$) and $H_{\Omega_{ccc}}^N$ ($H^{\Lambda}_{\Omega_{ccc}s}$) systems with the same total isospin share identical central terms. The spin-spin interaction terms depend on the flavor and spin wave functions of the considered systems. Since the meson-baryon $P_{\psi}^N/P^\Lambda_{\psi s}$ and di-baryon $H_{\Omega_{ccc}}^N/H^{\Lambda}_{\Omega_{ccc}s}$ systems have different spin wave functions, we can not directly relate the effective potentials from the $P_{\psi}^N/P^\Lambda_{\psi s}$ to the $H_{\Omega_{ccc}}^N/H^{\Lambda}_{\Omega_{ccc}s}$ systems. Instead, we simply collect six groups (A-F) of the $P_{\psi}^N/P^\Lambda_{\psi s}/H_{\Omega_{ccc}}^N/H^{\Lambda}_{\Omega_{ccc}s}$ systems that share identical effective potentials. From groups A-F, we find that with the same effective potentials, the $H_{\Omega_{ccc}}^N$/$H^{\Lambda}_{\Omega_{ccc}s}$ systems bind much deeper than that of the $P_{\psi}^N/P^\Lambda_{\psi s}$ systems due to their larger reduced masses.

Then we perform  multi-channel calculations to the $P_{\psi}^N$ and $H_{\Omega_{ccc}}^N$ systems. Since the corrections of the coupled-channel effect for the $P_{\psi}^N$ and $H_{\Omega_{ccc}}^N$ mass spectra are introduced from the off-diagonal spin-spin interactions, the coupled-channel effect give small corrections to the masses of the $P_{\psi}^N$ and $H_{\Omega_{ccc}}^N$ bound states. We present the $P_{\psi}^N$ and $H_{\Omega_{ccc}}^N$ mass spectra in both scenarios, the different arrangements of the bound state solutions in these two scenarios can be used to test the flavor-spin symmetry in the $P_{\psi}^N$ and $H_{\Omega_{ccc}}^N$ systems.

We also perform the multi-channel calculations to the $P^\Lambda_{\psi s}$ and $H^{\Lambda}_{\Omega_{ccc}s}$ systems. Since the couplings of the $\Lambda_c\bar{D}_s^{(*)}-\Xi_c\bar{D}^{(*)}$ and $\Lambda_c\Omega_{cc}^{(*)}-\Xi_c\Xi_{cc}^{(*)}$ are introduced through the exchanges of the strange scalar and axial-vector light mesons, we introduce an factor $g_x$ with its range in $[0,1]$ to quantify the SU(3) breaking effect. The effective potentials of the $\Lambda_c\bar{D}_s^{(*)}-\Xi_c\bar{D}^{(*)}$ and $\Lambda_c\Omega_{cc}^{(*)}-\Xi_c\Xi_{cc}^{(*)}$ channels have non-vanishing contributions from the central terms, thus, the masses of the $\Lambda_c\bar{D}_s^{(*)}$ ($\Lambda_c\Omega_{cc}^{(*)}$) and $\Xi_c\bar{D}^{(*)}$ ($\Xi_c\Xi_{cc}^{(*)}$) bound states have considerable corrections from the SU(3) breaking effect. By comparing the mass spectra of the $P_{\psi}^N$ ($H_{\Omega_{ccc}}^N$) with $P^\Lambda_{\psi s}$ ($H_{\Omega_{ccc}s}^{\Lambda}$) systems, we conclude that the bound states that related to the $\Lambda_c\bar{D}^{(*)}$, $\Xi_c\bar{D}^{(*)}$, $\Lambda_c\Omega_{cc}^{(*)}$, and $\Xi_c\Xi_{cc}^{(*)}$ channels do not have their $P_{\psi}^N$ and $H_{\Omega_{ccc}}^N$ molecular partners. The emergences of these states are due to the SU(3) breaking effect. For the rest of the molecular states in the $P^\Lambda_{\psi s}$ and $H^{\Lambda}_{\Omega_{ccc}s}$ states, they can find their corresponding $P_{\psi}^N$ and $H_{\Omega_{ccc}}^N$ molecular partners with lowest isospin and identical total angular momentum numbers, the mass arrangements of these states is very similar to that of the $P_{\psi}^N$ and $H_{\Omega_{ccc}}^N$ mass spectra, this is the manifestation of the SU(3) flavor symmetry. The mass arrangements of the $P^\Lambda_{\psi s}/H^{\Lambda}_{\Omega_{ccc}s}$ spectra can also be used to test the existence of flavor-spin symmetry.

By checking the binding energies of the systems in A-F groups obtained from the single-channel and multi-channel formalisms, we discuss the roles of the coupled-channel and SU(3) breaking effects on the violations of the flavor-spin symmetry. In the single-channel formalism, the flavor-spin symmetry works very well for the $P_{\psi}^N$ ($H_{\Omega_{ccc}}^N$) and $P^\Lambda_{\psi s}$ ($H^{\Lambda}_{\Omega_{ccc}s}$) systems. In the multi-channel formalism, the systems in groups A and F receive considerable corrections from both the SU(3) breaking and coupled-channel effects, while the systems in groups B, C, D, and E receive corrections from the coupled-channel effect, and the SU(3) breaking effect has tiny corrections to these systems. By comparing the results in groups B, C, D, and E, we find that the violations of the flavor-spin symmetry introduced from the coupled-channel effect is system-dependent, the flavor-spin symmetry in the systems of group C is violated, while it is still a good symmetry in the systems of groups B, D, and E. The similar binding energies from the $[\Sigma_c\bar{D}]_{\frac{1}{2}}$ and $[\Sigma_c^*\bar{D}]_{\frac{3}{2}}$ systems in group A, and the heavy flavor meson-baryon or di-baryon systems in groups B, D, E, is another manifestation of the flavor-spin symmetry. We hope that further investigations on these systems from both experiments and lattice QCD simulations could test our results.
\section*{Acknowledgments}
This work is supported by the National Natural Science Foundation of China under Grants No. 12305090, 12105072. B. W. is also supported by the Start-up Funds for Young Talents of Hebei University (No. 521100221021).


\begin{thebibliography}{300}
\bibitem{LHCb:2015yax}
R.~Aaij \textit{et al.} [LHCb],
\href{https://journals.aps.org/prl/abstract/10.1103/PhysRevLett.115.072001}{Phys. Rev. Lett. \textbf{115}, 072001 (2015)}.
\bibitem{LHCb:2019kea}
R.~Aaij \textit{et al.} [LHCb],
\href{https://journals.aps.org/prl/abstract/10.1103/PhysRevLett.122.222001}{Phys. Rev. Lett. \textbf{122}, no.22, 222001 (2019)}.
\bibitem{LHCb:2021chn}
R.~Aaij \textit{et al.} [LHCb],
\href{https://journals.aps.org/prl/abstract/10.1103/PhysRevLett.128.062001}{Phys. Rev. Lett. \textbf{128}, no.6, 062001 (2022)}.
\bibitem{LHCb:2022ogu}
R.~Aaij \textit{et al.} [LHCb],
\href{https://journals.aps.org/prl/abstract/10.1103/PhysRevLett.131.031901}{Phys. Rev. Lett. \textbf{131}, no.3, 031901 (2023)}.
\bibitem{LHCb:2020jpq}
R.~Aaij \textit{et al.} [LHCb],
\href{https://www.sciencedirect.com/science/article/pii/S2095927321001717?via%3Dihub}{Sci. Bull. \textbf{66}, 1278-1287 (2021)}.
\bibitem{Chen:2016qju}
H.~X.~Chen, W.~Chen, X.~Liu and S.~L.~Zhu,
\href{https://www.sciencedirect.com/science/article/pii/S037015731630103X?via%3Dihub}{Phys. Rept. \textbf{639}, 1-121 (2016)}.
\bibitem{Lebed:2016hpi}
R.~F.~Lebed, R.~E.~Mitchell and E.~S.~Swanson,
\href{https://www.sciencedirect.com/science/article/pii/S0146641016300734?via%3Dihub}{Prog. Part. Nucl. Phys. \textbf{93}, 143-194 (2017)}.

\bibitem{Esposito:2016noz}
A.~Esposito, A.~Pilloni and A.~D.~Polosa,
\href{https://www.sciencedirect.com/science/article/pii/S037015731630391X?via%3Dihub}{Phys. Rept. \textbf{668}, 1-97 (2017)}.
\bibitem{Hosaka:2016pey}
A.~Hosaka, T.~Iijima, K.~Miyabayashi, Y.~Sakai and S.~Yasui,
\href{https://academic.oup.com/ptep/article/2016/6/062C01/2240707?login=false}{PTEP \textbf{2016}, no.6, 062C01 (2016)}.
\bibitem{Guo:2017jvc}
F.~K.~Guo, C.~Hanhart, U.~G.~Mei\ss{}ner, Q.~Wang, Q.~Zhao and B.~S.~Zou,
\href{https://journals.aps.org/rmp/abstract/10.1103/RevModPhys.90.015004}{Rev. Mod. Phys. \textbf{90}, no.1, 015004 (2018)}
\href{https://journals.aps.org/rmp/abstract/10.1103/RevModPhys.94.029901}{[erratum: Rev. Mod. Phys. \textbf{94}, no.2, 029901 (2022)]}
\bibitem{Ali:2017jda}
A.~Ali, J.~S.~Lange and S.~Stone,
\href{https://www.sciencedirect.com/science/article/pii/S0146641017300716?via%3Dihub}{Prog. Part. Nucl. Phys. \textbf{97}, 123-198 (2017)}.
\bibitem{Liu:2019zoy}
Y.~R.~Liu, H.~X.~Chen, W.~Chen, X.~Liu and S.~L.~Zhu,
\href{https://linkinghub.elsevier.com/retrieve/pii/S0146641019300304}{Prog. Part. Nucl. Phys. \textbf{107}, 237-320 (2019)}.
\bibitem{Brambilla:2019esw}
N.~Brambilla, S.~Eidelman, C.~Hanhart, A.~Nefediev, C.~P.~Shen, C.~E.~Thomas, A.~Vairo and C.~Z.~Yuan,
\href{https://www.sciencedirect.com/science/article/pii/S0370157320301915?via%3Dihub}{Phys. Rept. \textbf{873}, 1-154 (2020)}
\bibitem{Lucha:2021mwx}
W.~Lucha, D.~Melikhov and H.~Sazdjian,
\href{https://www.sciencedirect.com/science/article/pii/S0146641021000211?via%3Dihub}{Prog. Part. Nucl. Phys. \textbf{120}, 103867 (2021)}.
\bibitem{Chen:2021ftn}
S.~Chen, Y.~Li, W.~Qian, Z.~Shen, Y.~Xie, Z.~Yang, L.~Zhang and Y.~Zhang,
\href{https://link.springer.com/article/10.1007/s11467-022-1247-1}{Front. Phys. \textbf{18}, 44601 (2023)}.
\bibitem{Chen:2022asf}
H.~X.~Chen, W.~Chen, X.~Liu, Y.~R.~Liu and S.~L.~Zhu,
\href{https://iopscience.iop.org/article/10.1088/1361-6633/aca3b6}{Rept. Prog. Phys. \textbf{86}, no.2, 026201 (2023)}.
\bibitem{Meng:2022ozq}
L.~Meng, B.~Wang, G.~J.~Wang and S.~L.~Zhu,
\href{https://www.sciencedirect.com/science/article/abs/pii/S0370157323001679?via%3Dihub}{Phys. Rept. \textbf{1019}, 1-149 (2023)}.
\bibitem{Savage:1990di}
M.~J.~Savage and M.~B.~Wise,
Phys. Lett. B \textbf{248}, 177-180 (1990)
\bibitem{Feijoo:2022rxf}
A.~Feijoo, W.~F.~Wang, C.~W.~Xiao, J.~J.~Wu, E.~Oset, J.~Nieves and B.~S.~Zou,
Phys. Lett. B \textbf{839}, 137760 (2023)
\bibitem{Yan:2022wuz}
M.~J.~Yan, F.~Z.~Peng, M.~S\'anchez S\'anchez and M.~Pavon Valderrama,
Phys. Rev. D \textbf{107}, no.7, 074025 (2023)
\bibitem{Nakamura:2022gtu}
S.~X.~Nakamura and J.~J.~Wu,
Phys. Rev. D \textbf{108}, no.1, L011501 (2023)
doi:10.1103/PhysRevD.108.L011501
[arXiv:2208.11995 [hep-ph]].
\bibitem{Giachino:2022pws}
A.~Giachino, A.~Hosaka, E.~Santopinto, S.~Takeuchi, M.~Takizawa and Y.~Yamaguchi,
Phys. Rev. D \textbf{108}, no.7, 074012 (2023)
\bibitem{Zhu:2022wpi}
J.~T.~Zhu, S.~Y.~Kong and J.~He,
Phys. Rev. D \textbf{107}, no.3, 034029 (2023)
\bibitem{Chen:2022wkh}
K.~Chen, Z.~Y.~Lin and S.~L.~Zhu,
Phys. Rev. D \textbf{106}, no.11, 116017 (2022)


\bibitem{LHCb:2017iph}
R.~Aaij \textit{et al.} [LHCb],
\href{https://journals.aps.org/prl/abstract/10.1103/PhysRevLett.119.112001}{Phys. Rev. Lett. \textbf{119}, no.11, 112001 (2017)}.
\bibitem{LHCb:2022rpd}
R.~Aaij \textit{et al.} [LHCb],
\href{https://link.springer.com/article/10.1007/JHEP05(2022)038}{JHEP \textbf{05}, 038 (2022)}.
\bibitem{Roberts:2007ni}
W.~Roberts and M.~Pervin,
\href{https://www.worldscientific.com/doi/abs/10.1142/S0217751X08041219}{Int. J. Mod. Phys. A \textbf{23}, 2817-2860 (2008)}.
\bibitem{Migura:2006ep}
S.~Migura, D.~Merten, B.~Metsch and H.~R.~Petry,
\href{https://link.springer.com/article/10.1140/epja/i2006-10017-9}{Eur. Phys. J. A \textbf{28}, 41 (2006)}.
\bibitem{Ebert:1996ec}
D.~Ebert, R.~N.~Faustov, V.~O.~Galkin, A.~P.~Martynenko and V.~A.~Saleev,
\href{https://link.springer.com/article/10.1007/s002880050534}{Z. Phys. C \textbf{76}, 111-115 (1997)}.
\bibitem{Gerasyuta:1999pc}
S.~M.~Gerasyuta and D.~V.~Ivanov,
\href{https://link.springer.com/article/10.1007/BF03035848}{Nuovo Cim. A \textbf{112}, 261-276 (1999)}.
\bibitem{Itoh:2000um}
C.~Itoh, T.~Minamikawa, K.~Miura and T.~Watanabe,
\href{https://journals.aps.org/prd/abstract/10.1103/PhysRevD.61.057502}{Phys. Rev. D \textbf{61}, 057502 (2000)}.
\bibitem{Wang:2021rjk}
J.~B.~Wang, G.~Li, C.~R.~Deng, C.~S.~An and J.~J.~Xie,
\href{https://journals.aps.org/prd/abstract/10.1103/PhysRevD.104.094008}{Phys. Rev. D \textbf{104}, no.9, 094008 (2021)}.

\bibitem{Ebert:2002ig}
D.~Ebert, R.~N.~Faustov, V.~O.~Galkin and A.~P.~Martynenko,
\href{https://journals.aps.org/prd/abstract/10.1103/PhysRevD.66.014008}{Phys. Rev. D \textbf{66}, 014008 (2002)}

\bibitem{Fleck:1989mb}
S.~Fleck and J.~M.~Richard,
\href{https://academic.oup.com/ptp/article/82/4/760/1823785}{Prog. Theor. Phys. \textbf{82}, 760-774 (1989)}.
\bibitem{Ponce:1978gk}
W.~Ponce,
\href{https://journals.aps.org/prd/abstract/10.1103/PhysRevD.19.2197}{Phys. Rev. D \textbf{19}, 2197 (1979)}.
\bibitem{He:2004px}
D.~H.~He, K.~Qian, Y.~B.~Ding, X.~Q.~Li and P.~N.~Shen,
\href{https://journals.aps.org/prd/abstract/10.1103/PhysRevD.70.094004}{Phys. Rev. D \textbf{70}, 094004 (2004)}.
\bibitem{Weng:2010rb}
M.~H.~Weng, X.~H.~Guo and A.~W.~Thomas,
\href{https://journals.aps.org/prd/abstract/10.1103/PhysRevD.83.056006}{Phys. Rev. D \textbf{83}, 056006 (2011)}.
\bibitem{Soto:2020xpm}
J.~Soto and J.~Tarr\'us Castell\`a,
\href{https://journals.aps.org/prd/abstract/10.1103/PhysRevD.102.014012}{Phys. Rev. D \textbf{102}, no.1, 014012 (2020)}.
\bibitem{Wei:2015gsa}
K.~W.~Wei, B.~Chen and X.~H.~Guo,
\href{https://journals.aps.org/prd/abstract/10.1103/PhysRevD.92.076008}{Phys. Rev. D \textbf{92}, no.7, 076008 (2015)}.
\bibitem{Oudichhya:2021yln}
J.~Oudichhya, K.~Gandhi and A.~K.~Rai,
\href{https://journals.aps.org/prd/abstract/10.1103/PhysRevD.104.114027}{Phys. Rev. D \textbf{104}, no.11, 114027 (2021)}.
\bibitem{Zhang:2008rt}
J.~R.~Zhang and M.~Q.~Huang,
\href{https://journals.aps.org/prd/abstract/10.1103/PhysRevD.78.094007}{Phys. Rev. D \textbf{78}, 094007 (2008)}.
\bibitem{Wang:2010it}
Z.~G.~Wang,
\href{https://link.springer.com/article/10.1140/epja/i2011-11081-8}{Eur. Phys. J. A \textbf{47}, 81 (2011)}.
\bibitem{Flynn:2003vz}
J.~M.~Flynn \textit{et al.} [UKQCD],
\href{https://iopscience.iop.org/article/10.1088/1126-6708/2003/07/066}{JHEP \textbf{07}, 066 (2003)}.
\bibitem{Perez-Rubio:2015zqb}
P.~P\'erez-Rubio, S.~Collins and G.~S.~Bali,
\href{https://journals.aps.org/prd/abstract/10.1103/PhysRevD.92.034504}{Phys. Rev. D \textbf{92}, no.3, 034504 (2015)}.
\bibitem{Bahtiyar:2020uuj}
H.~Bahtiyar, K.~U.~Can, G.~Erkol, P.~Gubler, M.~Oka and T.~T.~Takahashi,
\href{https://journals.aps.org/prd/abstract/10.1103/PhysRevD.102.054513}{Phys. Rev. D \textbf{102}, no.5, 054513 (2020)}.
\bibitem{Chen:2021cfl}
K.~Chen, R.~Chen, L.~Meng, B.~Wang and S.~L.~Zhu,
\href{https://link.springer.com/article/10.1140/epjc/s10052-022-10540-5}{Eur. Phys. J. C \textbf{82}, no.7, 581 (2022)}.
\bibitem{Chen:2021spf}
K.~Chen, B.~Wang and S.~L.~Zhu,
\href{https://journals.aps.org/prd/abstract/10.1103/PhysRevD.105.096004}{Phys. Rev. D \textbf{105}, no.9, 096004 (2022)}.
\bibitem{Wang:2023hpp}
B.~Wang, K.~Chen, L.~Meng and S.~L.~Zhu,
\href{https://journals.aps.org/prd/abstract/10.1103/PhysRevD.109.034027}{Phys. Rev. D \textbf{109}, no.3, 034027 (2024)}.
\bibitem{Wang:2023eng}
B.~Wang, K.~Chen, L.~Meng and S.~L.~Zhu,
\href{https://arxiv.org/abs/2312.13591}{[arXiv:2312.13591 [hep-ph]]}.
\bibitem{Guo:2013xga}
F.~K.~Guo, C.~Hidalgo-Duque, J.~Nieves and M.~P.~Valderrama,
\href{https://journals.aps.org/prd/abstract/10.1103/PhysRevD.88.054014}{Phys. Rev. D \textbf{88}, no.5, 054014 (2013)}.
\bibitem{Chen:2017jjn}
R.~Chen, A.~Hosaka and X.~Liu,
\href{https://journals.aps.org/prd/abstract/10.1103/PhysRevD.96.114030}{Phys. Rev. D \textbf{96}, no.11, 114030 (2017)}.
\bibitem{Asanuma:2023atv}
T.~Asanuma, Y.~Yamaguchi and M.~Harada,
\href{https://arxiv.org/abs/2311.04695}{[arXiv:2311.04695 [hep-ph]]}.
\bibitem{Junnarkar:2019equ}
P.~Junnarkar and N.~Mathur,
\href{https://journals.aps.org/prl/abstract/10.1103/PhysRevLett.123.162003}{Phys. Rev. Lett. \textbf{123}, no.16, 162003 (2019)}.
\bibitem{Pan:2019skd}
Y.~W.~Pan, M.~Z.~Liu, F.~Z.~Peng, M.~S\'anchez S\'anchez, L.~S.~Geng and M.~Pavon Valderrama,
\href{https://journals.aps.org/prd/abstract/10.1103/PhysRevD.102.011504}{Phys. Rev. D \textbf{102}, no.1, 011504 (2020)}.
\bibitem{Chen:2018pzd}
R.~Chen, F.~L.~Wang, A.~Hosaka and X.~Liu,
\href{https://journals.aps.org/prd/abstract/10.1103/PhysRevD.97.114011}{Phys. Rev. D \textbf{97}, no.11, 114011 (2018)}.

\bibitem{Leinweber:2003dg}
D.~B.~Leinweber, A.~W.~Thomas and R.~D.~Young,
\href{https://journals.aps.org/prl/abstract/10.1103/PhysRevLett.92.242002}{Phys.
Rev. Lett. \textbf{92}, 242002 (2004).}
\bibitem{Wang:2007iw}
P.~Wang, D.~B.~Leinweber, A.~W.~Thomas and R.~D.~Young,
\href{https://journals.aps.org/prd/abstract/10.1103/PhysRevD.75.073012}{Phys.
Rev. D \textbf{75}, 073012 (2007).}
\end{thebibliography}
\end{document}